\documentclass[prd,aps,twocolumn,a4paper,showkeys,nofootinbib,superscriptaddress]{revtex4-2}

\usepackage[normalem]{ulem}
\usepackage{graphicx}% Include figure files
\usepackage{dcolumn}% Align table columns on decimal point
\usepackage{bm}% bold math
\usepackage{hyperref}% add hypertext capabilities
\usepackage[mathlines]{lineno}% Enable numbering of text and display math
\usepackage{amsmath,amssymb,euscript}
\usepackage{calrsfs}
\allowdisplaybreaks
\usepackage{booktabs} % Per una migliore qualità delle tabelle
\usepackage{multirow}

\usepackage{tikz}
\usetikzlibrary{calc,decorations.markings}
\usetikzlibrary{snakes}
\usepackage{rotating}
\usepackage{comment}
\usepackage{empheq}
\usepackage{overpic}
\usepackage{hyperref}

\allowdisplaybreaks[2]
\numberwithin{equation}{section}	
\renewcommand{\theequation}{\arabic{section}.\arabic{equation}}

\definecolor{darkgreen}{rgb}{0.0, 0.55, 0.1}

\usepackage[dvipsnames]{xcolor}

\begin{document}

\title{Observable signature of  magnetic tidal coupling in hierarchical triple systems}

\author{Marta Cocco}
\email{marta.cocco@nbi.ku.dk}
 \affiliation{Dipartimento di Fisica e Geologia, Universit\`a di Perugia, I.N.F.N. Sezione di Perugia, \\ Via Pascoli, I-06123 Perugia, Italy}
 \affiliation{Center of Gravity, Niels Bohr Institute, Copenhagen University,\\ Blegdamsvej 17, DK-2100 Copenhagen \O{}, Denmark}
\author{Gianluca Grignani}
\email{gianluca.grignani@unipg.it}
\affiliation{Dipartimento di Fisica e Geologia, Universit\`a di Perugia, I.N.F.N. Sezione di Perugia, \\ Via Pascoli, I-06123 Perugia, Italy}
\author{Troels Harmark}
\email{harmark@nbi.ku.dk}
 \affiliation{Center of Gravity, Niels Bohr Institute, Copenhagen University,\\ Blegdamsvej 17, DK-2100 Copenhagen \O{}, Denmark}
  \affiliation{Nordita, KTH Royal Institute of Technology and Stockholm University,\\
  Hannes Alfv\'ens v\"ag 12, SE-106 91 Stockholm, Sweden}

\author{Marta Orselli}
\email{marta.orselli@unipg.it}
 \affiliation{Dipartimento di Fisica e Geologia, Universit\`a di Perugia, I.N.F.N. Sezione di Perugia, \\ Via Pascoli, I-06123 Perugia, Italy}
 \affiliation{Center of Gravity, Niels Bohr Institute, Copenhagen University,\\ Blegdamsvej 17, DK-2100 Copenhagen \O{}, Denmark}
\author{Davide Panella}
\email{davide.panella@nbi.ku.dk}
  \affiliation{Dipartimento di Fisica e Geologia, Universit\`a di Perugia, I.N.F.N. Sezione di Perugia, \\ Via Pascoli, I-06123 Perugia, Italy}
 \affiliation{Center of Gravity, Niels Bohr Institute, Copenhagen University,\\ Blegdamsvej 17, DK-2100 Copenhagen \O{}, Denmark}
\author{Daniele Pica}
\email{daniele.pica@nbi.ku.dk}
 \affiliation{Dipartimento di Fisica e Geologia, Universit\`a di Perugia, I.N.F.N. Sezione di Perugia, \\ Via Pascoli, I-06123 Perugia, Italy}
 \affiliation{Institut f\"ur Theoretische Physik, Goethe Universit\"at,
 \\ Max-von-Laue-Str. 1, 60438 Frankfurt am Main, Germany \\ (Authors appear in alphabetical order)}

\begin{abstract}
We study hierarchical triple systems formed by a compact binary orbiting a supermassive black hole (SMBH), focusing on the role of relativistic magnetic tidal interactions. Extending previous analyses of precession resonances to 0.5 post-Newtonian order, we incorporate quadrupolar magnetic tidal moments, which have no Newtonian counterpart. We find that magnetic tides introduce new resonances 
absent at lower order, leading to additional eccentricity excitations and significantly modifying the binary’s long-term evolution. Numerical solutions of the Lagrange Planetary Equations confirm these analytical predictions and reveal how resonance strength depends on orbital eccentricity and inclination. The resulting dynamics accelerates the binary merger and imprints distinctive signatures on gravitational waves, potentially observable by LISA. Our findings identify magnetic tidal coupling as a novel strong-gravity effect and establish its importance for the resonant dynamics of compact-object binaries near SMBHs.

\end{abstract}

\maketitle

\tableofcontents

\section{Introduction}

Tidal forces play a central role in the dynamics of three-body systems.
In Newtonian gravity, tidal effects depend only on the positions and masses of the bodies, and they are crucial to understanding several phenomena.
For instance, ocean tides arise from tidal forces in the Earth-Moon-Sun system and, similarly, the orbital motion of satellites can be altered by the tidal perturbation of a third body \cite{Newton:1687eqk,euler1772theoria,Lagrange1783,delaunay1860théorie,Poincaré1891,Love}.

Extending these phenomena to the strong-gravity regime can reveal fundamental insights for future gravitational physics. For example, the tidal deformability of neutron stars can constrain their nuclear equation of state \cite{Hinderer:2007mb,Flanagan:2007ix,Damour:2009vw,Binnington:2009bb}, and both the secular and non-secular evolution of a coalescing compact binary can be affected by the tidal perturbation of a third body \cite{Samsing:2024syt,Hendriks:2024gpp, Saini:2025ncj}.
A well-known secular effect is the von Zeipel–Lidov–Kozai (ZLK) mechanism \cite{vonzeiepel,Kozai,Lidov}, which induces large oscillations in the binary’s eccentricity over long timescales \cite{Hu:2024hzu, Chandramouli_Yunes, Randall:2018qna,Camilloni_2024,Antonini:2012ad,Maeda_2023,maeda2023chaoticvonzeipellidovkozaioscillations,Maeda:2025row,Naoz_2016}. 
An example of non-secular dynamics involves resonances, which can imprint characteristic signatures on the gravitational waveform \cite{Bonga_et_al_2019,Gupta_et_al_2021,Flanagan_Hughes_Ruangsri,Flanagan_Hinderer_2012,Zwick:2025ine}, induce episodic eccentricity growth that accelerates the merger process \cite{Sharpe:2025voe,Bhaskar:2022tzq}, and potentially affect the long-term stability of the binary system \cite{Stockinger:2024tai,Stockinger:2025ouz}.

In a relativistic framework, the tidal environment is encoded in the spacetime metric, which can be decomposed into electric and magnetic tidal fields \cite{Thorne:1984mz,PoissonVlasov}.
The electric tidal fields generalize the Newtonian tidal force to the strong-gravity regime. Several studies have demonstrated that these relativistic tidal effects can leave observable imprints \cite{Camilloni_2024,Maeda_2023,maeda2023chaoticvonzeipellidovkozaioscillations,Maeda:2025row,Cocco:2025adu}.
General Relativity also predicts a qualitatively new phenomenon absent in Newtonian gravity: the magnetic tidal fields. These fields have no weak-field analogue and represent a new kind of velocity-dependent tidal interaction. Any observable signature arising from magnetic tidal fields would therefore constitute a direct test of General Relativity beyond the weak-field limit.

In this work, we identify new resonances induced by a magnetic tidal field, which can drive the eccentricity growth of a compact binary orbiting in the vicinity of a SMBH.
Such hierarchical triple systems, where a compact binary orbits close to an SMBH, are known as binary Extreme Mass Ratio Inspirals (b-EMRIs)~\cite{Generozov,Chen:2018axp}.
% {\sl i.e.}~a b-EMRI three-body system \cite{Samsing:2017rat, Samsing:2024syt, Generozov, Chen:2018axp}. 
This discovery reveals a potentially observable signature of magnetic tidal effects acting on compact binaries.

The new resonances that we find generalize the previously found {\it precession resonances} between the frequencies of the orbit around the SMBH and the precession frequency of the compact binary. These were first studied in \cite{Kuntz:2021hhm} using a post-Newtonian (PN) approach, 
in which the components of the triple system were treated as point particles and evolved under Newtonian ($0$PN) dynamics.
 This analysis was extended in Ref.~\cite{Cocco:2025adu} to include strong-gravity electric tidal fields, by retaining a $0$PN description for the compact binary while modeling its interaction with the SMBH relativistically through quadrupolar electric tidal moments~\cite{PoissonVlasov}, with the SMBH described by a Schwarzschild spacetime.
The main result of Ref.~\cite{Cocco:2025adu} is that incorporating strong-gravity effects on the electric tidal fields enriches the resonance spectrum, as the multiple fundamental frequencies of the outer motion in curved spacetime enter the resonance condition.
The resonance condition derived in Ref.~\cite{Cocco:2025adu} takes the form
\begin{equation} 
    \label{eq:intro}
    q\, \dot{\gamma} = k \,\Omega_{\hat{r}} + l \,\Omega_{\hat{\Psi}} \,,
\end{equation}
where $\dot\gamma$ is the precession frequency of the compact binary,
$\Omega_{\hat{r}}$ is the angular frequency associated with the radial motion of the outer orbit
and $\Omega_{\hat{\Psi}}$ is the angular frequency associated with the orientation of the local inertial frame of the binary system \cite{van_de_Meent_2020, Bini_2016,Schmidt_2002,Fujita_2009,Hinderer_2008}. For quadrupolar electric tidal moments, the corresponding mode index is $q=2$.

In this work, we incorporate a $0.5$PN-order coupling between the quadrupolar magnetic tidal moments induced by the SMBH and the compact binary.
This coupling gives rise to additional resonances beyond those produced by the quadrupolar electric tidal moments.
The odd-parity nature of the magnetic tidal field, reflected in its coupling to the binary dynamics, enforces $q = 1$ in the resonance condition~(Eq.~\eqref{eq:intro}).
The resulting magnetic resonances generate additional jumps in the eccentricity of the binary system, significantly altering its evolution relative to Ref.~\cite{Cocco:2025adu}.
These resonances therefore provide a distinct observational signature of the magnetic tidal coupling between the compact binary and the SMBH.

The magnetic tidal resonances identified in this work require a compact binary on a close orbit around a SMBH.
Such hierarchical triple systems are expected to form in dense galactic nuclei, which host SMBHs and a high density of stellar remnants, including compact-object binaries.
These binaries may arise through dynamical interactions~\cite{OLeary:2008myb,Samsing:2017rat} or be captured by the gravitational potential of the SMBH~\cite{Hopman:2006qr}.
In active galactic nuclei (AGNs), migration traps can further enhance the formation and long-term survival of tight compact-object binaries near the central SMBH~\cite{Bellovary:2015ifg,Secunda:2020mhd,Peng_Chen,Tagawa_2020}.

Additional environmental effects, such as those due to gas, stars, or other surrounding matter, can influence the dissipative evolution of a compact binary~\cite{Cardoso:2020iji, Zwick:2021ayf, CanevaSantoro:2023aol, Ishibashi:2024wwu, Barausse:2007dy, Barausse:2014tra, Cardoso:2020lxx, Cole:2022yzw, Chen:2025qyj}. In this work, we focus solely on the tidal forces generated by the background spacetime through which the binary moves, neglecting all other environmental contributions.  
%\sout{We expect, in fact, that such effects modify only the timing at which a resonance is crossed, without altering the number or the strength of the precession resonances.}
We find numerically, within the model adopted here (see App.~\ref{App:Environmental effects}), that such effects modify only the timing at which a resonance is crossed, without altering the number or the strength of the precession resonances.

The GW190521 event~\cite{LIGOScientific:2020iuh}, a merger of two stellar-mass black holes, has been interpreted as a candidate for having occurred within a migration trap in an AGN~\cite{Toubiana:2020drf, Graham:2020gwr, Sberna:2022qbn}. Similar migration processes may also account for the formation of a stellar-mass binary orbiting near Sgr A*, as recently reported in Ref.~\cite{Peissker:2024ade}. 
In addition, binaries passing close to an SMBH can be tidally captured, giving rise to b-EMRIs~\cite{Generozov, Chen:2018axp}.

Owing to their relevance as multi-band gravitational waves (GWs) sources and their potential for testing general relativity, b-EMRI systems have been investigated in a variety of scenarios, with particular attention to their dynamics and GW emission \cite{Santos:2025ass, Chen:2018axp, Camilloni_2024, Yin:2024nyz, Meng:2024kug, Jiang:2024mdl}.
Understanding the physics governing these systems is essential for future observations, both to mitigate biases in data analysis and to identify signatures of strong-gravity effects. This is particularly relevant for LISA for which EMRIs, and consequently b-EMRIs, are expected to constitute one of the primary sources~\cite{Amaro-Seoane:2017sbf}.

In this work, we show that magnetic tidal fields trigger precession resonances absent at $0$PN order, producing characteristic signatures entirely within the LISA frequency band. These resonances offer, to the best of our knowledge, the first potentially detectable probe of purely relativistic magnetic tidal coupling in the resonant dynamics of compact binaries.

The paper is organized as follows. In Sec.~\ref{sec:1/2PNsystem}, we present the analytical framework used to derive the system’s Hamiltonian up to $0.5$PN order.
The parametrization adopted to describe the system, based on an underlying gauge symmetry, is detailed in Sec.~\ref{sec:orbital_gauge}.
In Sec.~\ref{average} we perform the averaging of the Hamiltonian over the inner orbit. 
Section~\ref{sec:analytic} introduces the fundamental frequencies of the outer motion, and develops an analytical model for the resonances induced by the magnetic tidal interaction in the limit of small external eccentricity.
In Sec.~\ref{sec:Numerical} we present numerical solutions for the compact-binary evolution, validating the analytical predictions and extending them to scenarios with finite external eccentricity.
Sec.~\ref{sec:CONCLUSIONS} summarizes our results and discusses possible future extensions. 
In Appendix~\ref{app.tetrads}, we review the construction of Marck’s tetrad and introduce the {\it accelerated tetrad} used to decouple the center-of-mass and internal motions of the binary up to the $0.5$PN order.
Appendix~\ref{Tidal_Moments_Schwarzschild} provides the explicit expressions for the electric and magnetic tidal moments associated with a Schwarzschild geodesic, which form the basis for computing the tidal couplings entering the Hamiltonian.
In Appendix~\ref{App:DistantStarFrame} we introduce the {\it distant-star frame}, obtained through a time-dependent rotation of Marck’s tetrad, which allows us to connect the local inertial description to that of a distant (asymptotic) observer and to identify the contribution associated with {\it gyroscope precession}.
Finally, Appendix~\ref{App:Environmental effects} provides a qualitative overview of how other non-tidal environmental effects, neglected throughout the main text, could influence the evolution and resonant behavior of the binary system.

For clarity, when the same symbol is used for both the inner and outer motion, hatted quantities denote outer-orbit variables, while unhatted quantities refer to the inner binary. 
Greek indices run from $0$ to $3$ and Latin indices from $1$ to $3$, except for the mode indices $(k,l)$ in the resonance condition. 
We adopt the metric signature $(-,+,+,+)$.

\section{Hierarchical triple system up to 0.5PN order}
\label{sec:1/2PNsystem}
In this section, we review the basic dynamics of a binary system subject to tidal forces arising from the spacetime curvature generated by an external SMBH. 
The binary consists of two compact objects with masses $m_1$ and $m_2$, which may be either stellar-mass black holes or neutron stars; we refer to this as the \textit{inner binary}. 
The two compact objects are assumed to be in a bound motion referred to as the \textit{inner orbit}.
The center of mass (CoM) of the inner binary is itself in bound motion around the external SMBH of mass $M_*$, which we refer to as the \textit{outer orbit}.

To maintain analytical control over the dynamics, we adopt two simplifying assumptions:
\begin{itemize}

    \item We assume the \textit{small-tide approximation}~\cite{PoissonVlasov}, 
    which allows us to treat the interaction between the SMBH and the inner binary perturbatively in terms of the ratio between the binary’s size $r$ and the curvature radius $\mathcal{R}$ of the SMBH, i.e., $r/\mathcal{R} \ll 1$. We retain only the leading (quadrupolar) term, which scales as $\left(r/\mathcal{R}\right)^2$.
    The small-tide approximation is satisfied by assuming a hierarchical mass configuration of the triple system, $m_1,m_2 \ll M_*$, enabling us to capture strong relativistic effects in the tidal coupling between the SMBH and the inner binary~\cite{Camilloni_2024, Cocco:2025adu, Camilloni_2023}.
    \item We require the separation $r$ between the compact objects in the inner binary to be much larger than their Schwarzschild radii. This ensures that the dynamics can be modeled within Newtonian physics, supplemented with PN corrections to account for effects such as periastron precession and RR.

\end{itemize}

We focus on the long-term evolution of the inner binary, considering timescales longer than its orbital period but comparable to that of the outer orbit, as these are the timescales  
over which precession resonances can occur.
In this regime, the inner binary can be effectively modeled as an effective point particle moving in the gravitational field of the external SMBH, thereby reducing the problem to an effective two-body system~\cite{Kuntz_2021,Kuntz_Trincherini_Serra}.
We also neglect GW emission from the outer orbit, since it modifies the outer orbital parameters on a timescale much longer than both the characteristic evolution timescale of the inner binary and the outer orbital period~\cite{Hinderer_2008}.

In the small-tide approximation, the spacetime metric in the neighborhood of a reference geodesic can be expressed as the Minkowski metric $\eta_{\mu\nu}$ plus tidal perturbations~\cite{Camilloni_2024,Maeda_2023,maeda2023chaoticvonzeipellidovkozaioscillations}.
At leading (quadrupolar) order, these perturbations are characterized by the electric and magnetic tidal tensors, $\mathcal{E}_{ij}$ and $\mathcal{B}_{ij}$, which are defined as the frame components of the Weyl tensor $C_{\mu\nu\rho\sigma}$ evaluated along the geodesic~\cite{PoissonVlasov,Camilloni_2023}.
To construct these components, we introduce an orthonormal tetrad $\lambda_A{}^\mu$ (with $A = 0,1,2,3$) that defines a local reference frame at each point along the geodesic, without assuming any specific transport law such as Fermi–Walker transport.
The relevant components of the Weyl tensor in the tetrad frame are
\begin{equation} 
    \label{Weylcomp}
    \begin{aligned}
    C_{i j} & \equiv C_{\mu \nu \rho \sigma} \lambda_0^{\mu} \lambda_i^{\nu} \lambda_0^{\rho} \lambda_j^{\sigma} \,,\\
    C_{i j k} & \equiv C_{\mu \nu \rho \sigma} \lambda_i^{\mu} \lambda_j^{\nu} \lambda_k^{\rho} \lambda_0^{\sigma} \,,
\end{aligned}
\end{equation}
where indices $i,j,k = 1,2,3$ label spatial components. The electric and magnetic quadrupole tidal tensors are then defined as~\cite{PoissonVlasov}
\begin{equation} 
    \label{Tidaldef}
    \mathcal{E}_{i j} \equiv C_{i j}\,, \quad \mathcal{B}_{i j} \equiv \frac{1}{2} \epsilon_{ipq} C^{pq}{}_{j} \,,
\end{equation}
where $\epsilon_{ijk}$ is the three-dimensional Levi-Civita symbol with $\epsilon_{123} = +1$.
\footnote{Both $\mathcal{E}_{ij}$ and $\mathcal{B}_{ij}$ are symmetric and trace-free by construction~\cite{PoissonVlasov}.}

At quadrupolar order and up to $0.5$PN accuracy, the Lagrangian for each compact object $m_I$ ($I=1,2$) reads~\cite{Camilloni_2024}
\begin{equation}
\begin{aligned}
    L_{I}=& \left.L_{I}\right|_{\mathcal{E}=\mathcal{B}=0}-\frac{1}{2} m_I c^2 x_{I}^i x_{I}^j \mathcal{E}_{i j}\\
    &-\frac{2 c}{3} m_I v_{I}^i x_{I}^k x_{I}^l \epsilon_{i j k} \mathcal{B}^j{ }_l \,,
\end{aligned}
\end{equation}
where $x^i$ are the local coordinates of the generic local frame defined by the  tetrad $\lambda^\mu_A$.
The acceleration and angular velocity of this frame are given by 
\begin{equation}
\label{aiomegaigeneriche}
    a^i \equiv \lambda^i_\mu\frac{D\lambda^\mu_0}{d\hat{\tau}} \,, \qquad
    {\omega}^i\equiv-\frac{1}{2}{\epsilon^{ij}}_{k}{{\lambda}}^\mu_j\frac{D{{\lambda}}_\mu^k}{d\hat{\tau}} \,,
\end{equation}
where $\hat{\tau}$ is the proper time of the geodesic. \\
Consequently, the Lagrangian of the inner binary at $0.5$PN order takes the form
\begin{equation}
    \label{L0plusL1/2}
    L_{\rm inner}=\sum_{I=1}^2L_I= L_{\mathrm{0PN}}+L_{\rm 0.5 PN} \,,
\end{equation}
where
\begin{equation}
    \label{L_0}
    L_{\mathrm{0PN}} \equiv \frac{1}{2}\left(m_1 v_1^2+m_2v_2^2\right)+\frac{G m_1 m_2}{\left|\boldsymbol{x_1}-\boldsymbol{x_2}\right|}+L_a+L_\omega+L_{\mathcal{E}}
\end{equation}
represents the Lagrangian of the binary system at $0$PN order. Explicitly, we have
\begin{subequations}
\begin{align}
    \label{eq:L_a}
    L_a & =-\sum_{I=1}^2 m_I a_{k} x_I^{k} \,,\\
    L_\omega & =-\sum_{I=1}^2 m_I\left[\epsilon_{j k l} \omega^{l} x_I^{k}  v_I^{j}-\frac{1}{2}\left(\omega^2 x_I^2-\left(\boldsymbol{\omega} \cdot \boldsymbol{x_I}\right)^2\right)\right] \,, \\
    L_{{\mathcal{E}}} & =-\frac{c^{2}}{2} \sum_{I=1}^2 m_I x_I^{i} x_I^{j} \mathcal{E}_{ij} \,,
\end{align}
\end{subequations}
and
\begin{equation}
    \label{L_1/2}
    L_{\rm 0.5PN} = -\frac{2}{3} \sum_{I=1}^2 m_I c^2 x_I^{k} x_I^{l} \frac{v_I^{i}}{c} \epsilon_{ijk}\mathcal{B}^{j}{}_{l} \,.
\end{equation}
Note that $L_a$ represents the contribution from the inertial (fictitious) force experienced by an accelerating observer, while $L_\omega$ arises from the rotation of the frame, corresponding to the centrifugal and Coriolis forces.
Both terms vanish in an inertial frame.

We introduce the CoM coordinates and the relative coordinates of the inner binary, defined as
\begin{equation}
\begin{aligned} \label{coord.rel.}
    & M=m_1+m_2~, \quad \mu=\frac{m_1 m_2}{M}~, \\
    & \boldsymbol{X}=\frac{m_1 \boldsymbol{x}_1+m_2 \boldsymbol{x}_2}{M}~, \quad \boldsymbol{x}=\boldsymbol{x}_2-\boldsymbol{x}_1~, \quad r=|\boldsymbol{x}|~,\\
    & \boldsymbol{V}=\frac{m_1 \boldsymbol{v}_{1}+m_2 \boldsymbol{v}_{2}}{M}~, \quad \boldsymbol{v}=\boldsymbol{v}_{2}-\boldsymbol{v}_{1}~.
\end{aligned}
\end{equation}
We note that, at $0$PN order ($L_{0\rm PN}$), the motion of the CoM decouples from the relative motion.
Consequently, it is consistent to place the CoM of the inner binary on the reference geodesic, i.e., to set $\boldsymbol{X}=0$.
However, when the $0.5$PN term is included, this decoupling is no longer possible.
This becomes evident upon rewriting the $0.5$PN Lagrangian in Eq.~\eqref{L_1/2} in terms of the relative and CoM coordinates 
\begin{equation}
    \begin{aligned}
        L_{\rm 0.5 PN} =& L_{\rm 0.5\text{-CoM}}(\boldsymbol{X}, \boldsymbol{V}) + L_{\rm 0.5 \text{-rel}}(\boldsymbol{x}, \boldsymbol{v})\\
        &+ L_{\rm 0.5\text{-int}}(\boldsymbol{X}, \boldsymbol{V}, \boldsymbol{x}, \boldsymbol{v})
    \end{aligned}
\end{equation}
where
\begin{subequations}
\begin{align}
&L_{\rm 0.5\text{-CoM}}  = -\frac{2}{3}c^2 M \epsilon_{ijk} \mathcal{B}^{j}{}_{l} X^{k} X^{l} \frac{V^{i}}{c} \,, \\
&L_{\rm 0.5\text{-rel}} = -\frac{2}{3}c^2\mu \frac{m_1 - m_2}{M} \epsilon_{ijk} \mathcal{B}^{j}{}_{l} x^{k} x^{l} \frac{v^{i}}{c} \,, \\
\label{L1/2interactionterm} &L_{\rm 0.5\text{-int}} = -\frac{2}{3} c^2 \mu \epsilon_{ijk} \mathcal{B}^{j}{}_{l} \left[x^{k} x^{l} \frac{V^{i}}{c} 
+ \left(X^{k} x^{l} + x^{k} X^{l}\right) \frac{v^{i}}{c}\right] \,.
\end{align}
\end{subequations}
The interaction term $L_{\rm 0.5\text{-int}}$ explicitly couples the CoM and relative motion of the inner binary.
Therefore, the condition $\boldsymbol{X} = 0$, which describes geodesic motion of the CoM in the absence of acceleration ($\boldsymbol{a} = 0$), no longer holds at the $0.5$PN order.
However, as noted in Ref.~\cite{Maeda_2023} and reviewed in App.~\ref{app.tetrads}, the CoM and internal motion of the binary can be decoupled consistently up to the $0.5$PN order.
This is accomplished by adopting an accelerated reference frame in which the frame-acceleration term in Eq.~\eqref{eq:L_a},
\begin{equation}
\label{eq:L_CoM_a}
L_{a} = -M\, a_k X^k\, ,
\end{equation}
exactly cancels the $0.5$PN interaction term $L_{\rm 0.5\text{-int}}$ in Eq.~\eqref{L1/2interactionterm}, up to a total time derivative.

As discussed in App.~\ref{app.tetrads}, the frame acceleration required for the decoupling is at least of quadrupolar order, implying that $\boldsymbol{X}$ is also of the same order.
Consequently, the terms $L_a$, $L_{\rm 0.5\text{-CoM}}$, and $L_{\rm 0.5\text{-int}}$ contribute only beyond this order.
Since our analysis is limited to the dynamics of the inner binary up to and including the quadrupolar order in the small-tide expansion, we can safely neglect deviations of the CoM from the reference geodesic and deviations of the reference frame from an inertial one.
We therefore set $\boldsymbol{X}=0$, $a^i=0$, and $\omega^i=0$ consistently within the accuracy of our approximation.

The construction of a tetrad defining a local inertial frame was first presented by Marck in Ref.~\cite{Marck:1983}.
As discussed in App.~\ref{app.tetrads}, this tetrad introduces the angle $\hat{\Psi}$, referred to as {\it Marck’s angle}, which guarantees that the tetrad is parallel transported along the geodesic~\cite{Marck:1983,Camilloni_2023,Camilloni_2024}.
In this local inertial frame, the Lagrangian governing the relative motion of the inner binary, subject to the tidal field of the external SMBH, can be written, to quadrupolar order in the small-tide approximation and up to $0.5$PN order in the PN expansion, as
\begin{equation} 
    \label{Lbinaryupto1/2PN}
    \begin{aligned}
    L_{\text{inner}}=&\frac{1}{2} \mu v^2+\frac{G M \mu}{r}-\frac{1}{2} \mu c^2 x^i x^j \mathcal{E}_{ij}\\
    &-\frac{2}{3} \mu c^2 \frac{m_1-m_2}{M} \frac{v^i}{c} x^k x^l \epsilon_{i j k} \mathcal{B}^{j}{}_{l} \,.
    \end{aligned}
\end{equation}
Here, we have used the relative and CoM coordinates defined in Eq.~\eqref{coord.rel.}, together with the Lagrangian in Eq.~\eqref{L0plusL1/2}, noting that $\boldsymbol{X}=0$ can be consistently imposed at the order of approximation considered.

The conjugate momenta derived from Eq.~\eqref{Lbinaryupto1/2PN}, can be written as
\begin{equation}
    p^i= \mu v^i-\frac{2}{3} \mu c \frac{m_1-m_2}{M} x^k x^l \epsilon_{m j k} \mathcal{B}^j{ }_l \delta^{im} \,,
\end{equation}
where $\delta^{ij}$ denotes the Kronecker delta, used here to raise and lower spatial indices. 

Using these relations, we perform a Legendre transformation of the Lagrangian to obtain the corresponding Hamiltonian.
The Hamiltonian governing the relative motion of the inner binary, accurate to $0.5$PN order and to quadrupolar order in the small-tide expansion, is given by
\begin{equation}
    \label{Hbinary}
    H_{\text{inner}}=H_{0\rm PN}+H_{0.5\rm PN} \,,
\end{equation}
where
\begin{equation} 
\begin{aligned} 
    \label{eq:Hinner0PN&1/2PN}
     H_{0\rm PN}&=\frac{p^2}{2 \mu}-\frac{G M \mu}{r}+\frac{1}{2} \mu c^2 x^i x^j \mathcal{E}_{ij} \,,\\
     H_{0.5\rm PN}&=\frac{2}{3} c^2 \frac{m_1-m_2}{M} \frac{p^i}{c} x^k x^l \epsilon_{i j k} \mathcal{B}^{j}{}_{l} \,.
    \end{aligned}
\end{equation}
The first term in the Hamiltonian~\eqref{Hbinary} describes the $0$PN dynamics of the binary system in the presence of a strong-gravity electric tidal field generated by its orbit around the SMBH~\cite{Camilloni_2024, Cocco:2025adu, Maeda_2023, maeda2023chaoticvonzeipellidovkozaioscillations, Maeda:2025row}.
The term $H_{0.5\rm PN}$ instead represents the $0.5$PN contribution, which couples the binary to the strong-gravity magnetic tidal field induced by the SMBH.
To the best of our knowledge, this coupling has previously been considered only in the context of secular effects within triple systems in the weak-field regime~\cite{Kuntz_Trincherini_Serra}.
\footnote{The Hamiltonian of a binary system orbiting a SMBH up to $0.5$PN order was also derived in Refs.~\cite{Maeda:2025row, Maeda_2023, maeda2023chaoticvonzeipellidovkozaioscillations}. However, those analyses focused on the secular evolution of equal-mass binaries, for which the $0.5$PN contribution in Eq.~\eqref{Hbinary} vanishes.}

Following \cite{PoissonVlasov}, we introduce the radial unit vector $\Omega^a=x^a / r$, where $r=\sqrt{\delta_{a b} x^a x^b}$, to define the quantities
\begin{equation}
        \mathcal{E}^{\mathrm{q}}=\mathcal{E}_{c d} \Omega^c \Omega^d \,, \quad
        \mathcal{B}_i^{\rm q} = \epsilon_{i j k} \Omega^j \mathcal{B}^{j}{}_{l} \Omega^l \,.
\end{equation}

Using these relations, we can rewrite the Hamiltonian~\eqref{Hbinary} as
\begin{equation} 
    \label{HbinaryEqBq}
    \begin{aligned}
       H_{\text{inner}}&=\frac{p^2}{2 \mu}-\frac{G M \mu}{r}+\frac{1}{2} \mu c^2 r^2 \mathcal{E}^{\rm q} \\
       &+ \frac{2}{3} \mu c^2 \frac{m_1-m_2}{M} r^2 \frac{p^i}{c} \mathcal{B}^{{\rm q}}_{i} \,,
        \end{aligned}
\end{equation}
where the first three terms correspond to the Newtonian and electric-tidal contributions, while the last term represents the leading magnetic-tidal coupling at $0.5$PN order.

For completeness, we also derive the Hamiltonian as viewed by a distant (asymptotic) observer in App.~\ref{App:DistantStarFrame}.

\section{Orbital elements and gauge symmetry}\label{sec:orbital_gauge}

To analyze the dynamics and evolution of the inner binary, we adopt the standard parametrization of an elliptic Keplerian orbit, suitable for describing its Newtonian motion. In this section, we briefly review the main relations and definitions that will be used in the subsequent analysis.

The unperturbed two-body problem is governed by~\cite{goldstein:mechanics, morbidelli, Valtonen_Karttunen_2006}
\begin{equation}
    \label{2bodyunperturbed}
    \ddot{\boldsymbol{x}} + \frac{G M}{x^3} \boldsymbol{x} = 0 \,,
\end{equation}
where $\boldsymbol{x}$ denotes the relative position vector.
The general solution to Eq.~\eqref{2bodyunperturbed} corresponds to a Keplerian ellipse for a bound orbit and can be expressed as
\begin{equation}
    \boldsymbol{x} = \boldsymbol{f}(C_1, \ldots, C_6, t) \,, \quad
    \dot{\boldsymbol{x}} = \boldsymbol{g}(C_1, \ldots, C_6, t) \,,
\end{equation}
where $\boldsymbol{g} = \left(\partial \boldsymbol{f}/\partial t\right)_{C = \mathrm{const}}$ and $C_i$ $(i = 1, \ldots, 6)$ are constants of motion that can be written in terms of the Keplerian orbital elements or alternative sets of canonical variables, such as the Delaunay variables~\cite{delaunay1860théorie}.
The relative position vector for an elliptic orbit can then be parametrized as
\begin{equation}
    \label{ruv}
    \boldsymbol{x}=r (\cos\psi~ \boldsymbol{\hat u}+\sin\psi ~\boldsymbol{\hat v}) \,,
\end{equation}
with 
\begin{equation}
    \label{eq: r_ea}
    r=\frac{a (1-e^2)}{1+e \cos\psi} \,.
\end{equation}
Here, we introduce the \textit{semi-major axis} $a$, the \textit{eccentricity} $e$, and the \textit{true anomaly} $\psi$.
The unit vectors $\boldsymbol{\hat u}$ and $\boldsymbol{\hat v}$ denote, respectively, the \textit{periastron direction} and the \textit{direction of the ascending node}, forming an orthonormal triad together with the angular momentum direction, $\boldsymbol{\hat L}_{\rm in} = \boldsymbol{\hat u} \times \boldsymbol{\hat v}$.
For completeness, we also recall that, in a $0$PN two-body system, the \textit{eccentric anomaly} $\zeta$ and the \textit{mean anomaly} $\beta$ are defined as
\begin{equation}
    \label{beta_psi}
    \cos\psi=\frac{\cos\zeta-e}{1-e \cos\zeta}\,, \quad \beta=\zeta-e\sin \zeta \,.
\end{equation}

To characterize the orientation of the inner binary’s orbit, we introduce the standard Euler angles $\left(\vartheta, I, \gamma\right)$, where $\vartheta$ is the \textit{longitude of the ascending node}, $I$ is the \textit{inclination} of the inner orbit relative to the reference plane (taken here to be the plane of the outer orbit), and $\gamma$ is the \textit{argument of periastron}.
These angles can be associated with the following rotation matrices:
\begin{equation}
\begin{gathered}
R_\vartheta=\left(\begin{array}{ccc}
\cos \vartheta & -\sin \vartheta & 0 \\
\sin \vartheta & \cos \vartheta & 0 \\
0 & 0 & 1
\end{array}\right)\,, ~
R_I=\left(\begin{array}{ccc}
1 & 0 & 0 \\
0 & \cos I & -\sin I \\
0 & \sin I & \cos I
\end{array}\right)\,, ~
\\R_\gamma=\left(\begin{array}{ccc}
\cos \gamma & -\sin \gamma & 0 \\
\sin \gamma & \cos \gamma & 0 \\
0 & 0 & 1
\end{array}\right)\,,
\end{gathered}
\end{equation}
which define the general orientation of the inner orbit as
\begin{equation}
    \boldsymbol{x}=R_\vartheta R_I R_\gamma(r \cos \psi ~ \hat{\boldsymbol{x}}+r \sin \psi ~ \hat{\boldsymbol{y}}) \, .
\end{equation}
Consistent with Eq.~\eqref{ruv}, the unit vectors $\boldsymbol{\hat u}$ and $\boldsymbol{\hat v}$ are defined as
\begin{equation} 
    \label{eq:uandv}
    \boldsymbol{\hat u}=R_\vartheta R_I R_\gamma ~ \hat{\boldsymbol{x}} \,,\quad
    \boldsymbol{\hat v}=R_\vartheta R_I R_\gamma ~ \hat{\boldsymbol{y}} \,.
\end{equation}
The set of six parameters $\left(a, e, \beta, \vartheta, I, \gamma\right)$ thus fully specifies the orbital configuration of the inner binary.

Since the motion of the inner binary is periodic, it is convenient, following standard practice in celestial mechanics, to adopt the action–angle formalism to describe its dynamics.
To this end, we introduce the {\it Delaunay variables}~\cite{goldstein:mechanics, morbidelli, Valtonen_Karttunen_2006, delaunay1860théorie}, consisting of the three angle variables $\left(\beta, \gamma, \vartheta\right)$ and their conjugate action variables $\left(J_{\beta}, J_{\gamma}, J_{\vartheta}\right)$, defined as
\begin{equation} \label{actionvariables}
     J_\beta=\mu \sqrt{G M a} \,, \quad J_\gamma=J_\beta \sqrt{1-e^2}~, \quad J_{\vartheta}=J_\gamma \cos I \,.
\end{equation}
Note that $J_{\gamma}$ represents the magnitude of the inner binary’s orbital angular momentum, while $J_{\vartheta}$ corresponds to its projection onto the direction perpendicular to the reference plane, i.e.,
\begin{equation}
    J_\gamma=\left|\boldsymbol{L}_{\mathrm{in}}\right| \,,\quad
    J_{\vartheta}=\boldsymbol{L}_{\mathrm{in}} \cdot \hat{\boldsymbol{L}}_{\mathrm{out}} \,.
\end{equation}
where $\hat{\boldsymbol{L}}_{\mathrm{out}}$ is the unit vector along the angular momentum of the outer orbit, defining the normal to the reference plane.

So far, we have reviewed the dynamics of an unperturbed $0$PN two-body system.
In our setup, however, the inner binary is subject to the tidal field of the external SMBH and includes PN corrections up to $0.5$PN order.
To describe this perturbed two-body system, we introduce a source term $\boldsymbol{\mathcal{F}}$ in Eq.~\eqref{2bodyunperturbed}, accounting for the tidal interaction with the SMBH.
The resulting equation of motion reads
\begin{equation}
\label{2bodyperturbed}
    \ddot{\boldsymbol{x}}+\frac{GM}{x^3} \boldsymbol{x}=\boldsymbol{\mathcal{F}} \,.
\end{equation}
The solution to Eq.~\eqref{2bodyperturbed} with a source term can be expressed as
\begin{equation} 
    \label{rperturbed}
    \boldsymbol{x}=\boldsymbol{f}\left(C_1(t), \ldots, C_6(t), t\right) \,,
\end{equation}
where the six quantities $C_i(t)$ are now time-dependent orbital parameters.
The corresponding velocity is then given by
\begin{equation} 
    \label{velocityperturbed}
    \dot{\boldsymbol{x}}=\frac{\partial \boldsymbol{f}}{\partial t}+\sum_i \frac{\partial \boldsymbol{f}}{\partial C_i} \frac{\mathrm{~d} C_i}{\mathrm{~d} t}=\boldsymbol{g}+\sum_i \frac{\partial \boldsymbol{f}}{\partial C_i} \frac{\mathrm{~d} C_i}{\mathrm{~d} t}\,,
\end{equation}
Substituting the solution~\eqref{rperturbed} into Eq.~\eqref{2bodyperturbed} yields a system of three scalar equations for the six unknown functions $C_i(t)$, which makes the perturbed two-body problem underdetermined. This degeneracy can be lifted by imposing three additional independent constraints, which can be expressed as~\cite{gaugesymmetry}
\begin{equation} 
\label{generalgauge}
    \sum_i \frac{\partial \boldsymbol{f}}{\partial C_i} \frac{\mathrm{~d} C_i}{\mathrm{~d} t}=\boldsymbol{\Phi}\left(C_1, \ldots, C_6, t\right) \,,
\end{equation}
where $\boldsymbol{\Phi}$ is an arbitrary function of time. A different choice $\widetilde{\boldsymbol{\Phi}}$ leads to a corresponding set of functions $\widetilde{C}_i(t)$, yet both describe the same physical trajectory $\boldsymbol{x}(t)$.
This invariance under a change of parametrization defines a \textit{gauge symmetry} of the osculating elements~\cite{gaugesymmetry}. 

To derive the Lagrange Planetary Equations (LPE), which govern the time evolution of the parameters $C_i(t)$, one typically adopts the \textit{Lagrange gauge}, $\boldsymbol{\Phi} = 0$~\cite{Lagrange1808, Lagrange1809, Lagrange1810}. 
Nevertheless, the LPE can be derived for an arbitrary gauge choice satisfying Eq.~\eqref{generalgauge}.
In this more general case, the LPE expressed in terms of the Delaunay variables form a symplectic system only when the {\it generalized Lagrange gauge} is imposed~\cite{gaugesymmetry, EfroimskyGoldreichHJapproach}.
\begin{equation} 
    \label{GLG}
    \boldsymbol{\Phi}=-\frac{\partial \Delta \mathcal{L}}{\partial \dot{\boldsymbol{x}}}\,,
\end{equation}
where, in our context, $\Delta \mathcal{L}$ denotes the perturbative contributions to the inner binary’s Lagrangian arising from the tidal interaction with the external SMBH.

A key consequence of working in the generalized Lagrange gauge is that it modifies the definition of the momentum.
Combining Eqs.~\eqref{velocityperturbed} and~\eqref{generalgauge}, the momentum per unit mass can be written as
\begin{equation}
    \boldsymbol{p}=\boldsymbol{g}+\boldsymbol{\Phi}+\frac{\partial \Delta \mathcal{L}}{\partial \dot{\boldsymbol{x}}}\,,
\end{equation}
which, in the generalized Lagrange gauge defined by Eq.~\eqref{GLG}, simplifies to
\begin{equation}
    \boldsymbol{p}=\boldsymbol{g}\,.
\end{equation}
Therefore, by fixing the generalized Lagrange gauge, we can employ the canonical momentum of the unperturbed two-body system~\cite{gaugesymmetry} even in the presence of perturbations, namely, without explicitly including the tidal interaction with the SMBH in $\boldsymbol{p}$.
\footnote{If the perturbation to the Lagrangian does not depend on the velocities, the generalized Lagrange gauge condition reduces to $\boldsymbol{\Phi} = 0$, corresponding to the standard Lagrange gauge.
This means that in cases involving purely velocity-independent perturbations, such as those studied in~\cite{Camilloni_2024, Cocco:2025adu}, the generalized Lagrange gauge is effectively adopted ``implicitly''.}

It is worth noting that the ansatz introduced in Eq.~\eqref{rperturbed} can be applied in any chosen reference frame.
Consequently, the generalized Lagrange gauge condition~\eqref{GLG} can likewise be imposed in any frame, provided that the corresponding perturbation $\Delta \mathcal{L}$ is evaluated in that frame.

\section{Averaging over the inner orbit} \label{average}

Precession resonances can occur when the frequency of the inner binary periastron precession  becomes commensurate with the characteristic frequencies of the outer orbit.
Since the periastron precession timescale is significantly longer than the period of the inner orbit, we can study such resonances using the Hamiltonian \eqref{Hbinary} averaged over the inner orbit.

\subsection{Performing the average}
\label{subsec:validity_averaging}

Building on the discussion in Sec.~\ref{sec:orbital_gauge}, we express the Hamiltonian of the inner binary up to $0.5$ PN order in the generalized Lagrange gauge, using the following parametrization for the coordinates and momenta:
\begin{equation} 
    \label{kuntzparametrization}
    \begin{aligned}
        & \boldsymbol{x}=a \frac{\left(1-e^2\right)}{(1+e \cos{\psi})}(\cos{\psi}~\boldsymbol{\hat{u}} +\sin{\psi}~\boldsymbol{\hat{v}}) \,, \\
        & \boldsymbol{p}=\mu \sqrt{\frac{G M}{a (1-e^2)}} (-\sin{\psi}~\boldsymbol{\hat{u}}+(e+\cos{\psi})~\boldsymbol{\hat{v}}) \,.
    \end{aligned}
\end{equation}
Using Eq.~\eqref{beta_psi}, this parametrization reduces to the one employed in Ref.~\cite{Kuntz_Trincherini_Serra}.

To average over the inner orbit, we require a periodic angle that grows uniformly with time during the elliptic motion. In our setup, this role is fulfilled by the mean anomaly $\beta$. 
Since Eq.~\eqref{kuntzparametrization} depends on the true anomaly $\psi$, we compute the average as follows:
\begin{equation} 
    \label{averageintegral}
   \left<\mathcal{A}\right>=\frac{1}{2\pi}\int_0^{2\pi}\mathcal{A}~ d\beta= \frac{1}{2 \pi} \int_0^{2 \pi} \mathcal{A} \frac{\left(1-e^2\right)^{3 / 2}}{(1+e \cos \psi)^2}~d \psi \,.
\end{equation}
This defines the average over one inner orbital cycle, to leading order in the PN expansion.
This prescription is sufficient for our purposes, as Eq.~\eqref{averageintegral} remains valid even when the inner binary is modeled up to $0.5$PN order.
At higher PN orders, however, a modified averaging scheme must be adopted~\cite{Kuntz_2021, Kuntz_Trincherini_Serra}.
In the following, we apply this averaging procedure to the Hamiltonian in Eq.~\eqref{Hbinary}, which governs the inner-binary dynamics up to $0.5$PN order in the local inertial frame.
The Hamiltonian can be decomposed as
\begin{equation}
    \label{eq:H3pieces}
    H_{\text{inner}}=H|_{\mathcal{E}=\mathcal{B}=0}+H_{\mathcal{E}}+H_{\mathcal{B}} \,,
\end{equation}
where the first term
\begin{equation}
    H|_{\mathcal{E}=\mathcal{B}=0}=\frac{p^2}{2 \mu}-\frac{G M \mu}{r} \,,
\end{equation}
describes the dynamics of an isolated binary system. The remaining terms, 
\begin{equation}
    \begin{split}
      H_{\mathcal{E}}&=\frac{1}{2} \mu c^2 x^i x^j \mathcal{E}_{ij} \,,\\
         H_{\mathcal{B}}&=\frac{2}{3} c^2 \frac{m_1-m_2}{M} \frac{p^i}{c} x^k x^l \epsilon_{i j k} \mathcal{B}^{j}{}_{l} \,, 
    \end{split}
\end{equation}
represent the tidal interaction between the inner binary and the SMBH at $0$PN and $0.5$PN order, respectively, expressed in terms of the quadrupolar electric ($\mathcal{E}_{ij}$) and magnetic ($\mathcal{B}_{ij}$) tidal tensors. 

Using Eqs.~\eqref{kuntzparametrization} together with Eqs.~\eqref{actionvariables}, $H|_{\mathcal{E}=\mathcal{B}=0}$
can be written as
\begin{equation}
   H|_{\mathcal{E}=\mathcal{B}=0}=-\frac{G M \mu}{2 a} \,.
\end{equation} 
This term is independent of $\psi$ and therefore does not require averaging.

To perform the averaging over the inner orbit for the tidal part of the Hamiltonian in Eq.~\eqref{eq:H3pieces}, it is convenient to express the $0$PN quadrupolar electric contribution as~\cite{Cocco:2025adu}
\begin{equation}
    \label{electric}
    H_\mathcal{E}=\frac{1}{2} c^2  \mathcal{Q}^{ij} \mathcal{E}_{ij} \,,
\end{equation}
where $\mathcal{Q}^{ij}$ is given by
\begin{equation}
    \mathcal{Q}^{ij}=\mu x^i x^j \,.
\end{equation}
Using the parametrization \eqref{kuntzparametrization}, $\mathcal{Q}^{ij}$ can be expressed as
\begin{equation}
    \mathcal{Q}^{ij}=Q_1 u^i u^j+Q_2 v^i v^j+Q_3 (u^i v^j + v^i u^j) \,,
\end{equation}
where the dependence on the true anomaly $\psi$ is contained in the coefficients $Q_i$ for $i=1,2,3$.

Averaging over $\psi$ using the prescription in Eq.~\eqref{averageintegral} yields the orbit-averaged quadrupole electric tidal Hamiltonian:
\begin{equation}
\label{eq:secEpart}
           \left\langle H_\mathcal{E}\right\rangle=\frac{1}{2} c^2 \left\langle \mathcal{Q}^{ij}\right\rangle \mathcal{E}_{ij} \,,
\end{equation}
where the only non-vanishing contributions are 
\begin{equation} 
    \label{eq:averageQi}
    \begin{aligned}
        \left\langle Q_1\right\rangle=&\frac{1}{2} \mu a^2\left(1+4 e^2\right)~,\quad
        \left\langle Q_2\right\rangle=\frac{1}{2}\mu a^2\left(1-e^2\right) \,.\\
    \end{aligned}
\end{equation}
We now turn to averaging the $0.5$PN quadrupole magnetic contribution, $H_{\mathcal{B}}$.
Following the same reasoning used for the quadrupole electric term, we rewrite $H_{\mathcal{B}}$ as
\begin{equation} \label{HB}
    H_\mathcal{B}=\frac{2}{3} c ~ \mathcal{T}^{ij} \mathcal{B}_{ij} \,,
\end{equation}
where $\mathcal{T}^{ij}$ encodes the inner binary dynamics and is defined by
\begin{equation}
\mathcal{T}^{ij}=\frac{m_1-m_2}{M} p^{k} x^{l} x^{i} \epsilon_{k n l} \delta^{j n}~~.
\end{equation}
Using the parametrization in Eq.~\eqref{kuntzparametrization} and the symmetry properties of $\epsilon_{ijk}$, we obtain
\begin{equation}
\label{eq:Tij}
    \begin{aligned}
        \mathcal{T}^{ij}&=  \left(T_1 v^k u^l u^i-T_2 u^k v^l v^i\right) \epsilon_{k n l} \delta^{j n} \\
        &=\left(T_1 u^i+T_2 v^i\right)(\boldsymbol{\hat u} \times \boldsymbol{\hat v})^j \,,
    \end{aligned}
\end{equation}
where $T_1$ and $T_2$ encode the dependence on the true anomaly $\psi$. 
Averaging over the inner orbit then yields
\begin{equation}
\label{eq:secB}
    \left\langle H_\mathcal{B}\right\rangle=\frac{2}{3} c \left\langle\mathcal{T}^{ij}\right\rangle \mathcal{B}_{ij} \,,
\end{equation}
where the only nonzero contribution is 
\begin{equation} 
    \label{eq:averageTi}
    \left\langle T_1\right\rangle=-\frac{3}{2}\frac{m_1-m_2}{M}\mu  e a \sqrt{G M a(1-e^2)} \,.
\end{equation}

Therefore, the Hamiltonian in Eq.~\eqref{eq:H3pieces}, averaged over one inner orbit in a local inertial frame, is given by
\begin{equation} 
    \label{Hmarckaveinner}
    \left\langle H_{\text{inner}}\right\rangle=-\frac{G M \mu}{2 a}+\left\langle H_\mathcal{E}\right\rangle+\left\langle H_\mathcal{B}\right\rangle \,,
\end{equation}
where $\left<H_{\mathcal{E}}\right>$ and $\left<H_{\mathcal{B}}\right>$ are defined in Eqs.~\eqref{eq:secEpart}-\eqref{eq:averageQi} and~\eqref{eq:secB}-\eqref{eq:averageTi}, respectively.

Up to this point, we have not specified the explicit form of the tidal moments $\mathcal{E}_{ij}$ and $\mathcal{B}_{ij}$, which encode the influence of the external gravitational environment.
Consequently, the expression in Eq.~\eqref{Hmarckaveinner} remains valid for any background spacetime in which the inner binary is embedded.
From this point onward, in order to investigate precession resonances in the inner binary, we specialize to the case where the external SMBH is described by the Schwarzschild metric
\begin{equation}
    \label{sch_metric}
    ds^2=-\left(1-\frac{\hat{r}_s}{\hat{r}}\right)c^2 d\hat{t}^2 + \left(1-\frac{\hat{r}_s}{\hat{r}}\right)^{-1}d\hat{r}^2+\hat{r}^2d\hat{\Omega}^2 \,,
\end{equation}
where $\hat{r}_s =2 G M_*/c^2$ denotes the Schwarzschild radius and 
$ d\hat{\Omega}^2=d\hat{\theta}^2+\sin^2{\hat{\theta}} d\hat{\phi}^2$ represents the metric of a unit two-sphere. 
The explicit expressions for the tidal moments $\mathcal{E}_{ij}$ and $\mathcal{B}_{ij}$ generated by a Schwarzschild black hole are provided in App.~\ref{Tidal_Moments_Schwarzschild}.
Substituting these expressions into Eqs.~\eqref{eq:secEpart} and~\eqref{eq:secB} yields
\begin{equation}
    \begin{aligned}
    \label{eq:HEmarckaveinner}
    \left\langle  H_\mathcal{E}\right\rangle & =\mu c^2 \frac{\hat{r}_s a^2}{4 \hat{r}^3}\left[\frac{2+3 e^2}{2}\right. \\
    & +3 \frac{\hat{L}^2}{\hat{r}^2} \frac{2+3 e^2-5 e^2 \cos 2 \gamma}{4} \sin ^2 I \\
    & -3\left(1+\frac{\hat{L}^2}{\hat{r}^2}\right)\left(\frac{2+3 e^2+5 e^2 \cos 2 \gamma}{4} \cos ^2(\hat{\Psi}-\vartheta)\right. \\
    & +\frac{2+3 e^2-5 e^2 \cos 2 \gamma}{4} \sin ^2(\hat{\Psi}-\vartheta) \cos ^2 I \\
    & \left.\left.+\frac{5 e^2}{2} \sin 2 \gamma \cos (\hat{\Psi}-\vartheta) \sin (\hat{\Psi}-\vartheta) \cos I\right)\right] \,,
\end{aligned}
\end{equation}
and
\begin{equation} 
    \label{HBmarckaveinner}
    \begin{aligned}
     \left\langle H_\mathcal{B}\right\rangle&= \mu c \frac{m_1-m_2}{M}\frac{3 \hat{r}_s e a \hat{L}\sqrt{G M a (1-e^2)(\hat{L}^2+\hat{r}^2)}}{2 \hat{r}^5} \\
     &\times [\cos\gamma \cos I \cos(\vartheta-\hat{\Psi})-\sin\gamma \cos 2 I \sin (\vartheta-\hat{\Psi})] \,.
\end{aligned}
\end{equation}
We note that the averaged quadrupolar electric tidal term in Eq.~\eqref{eq:secEpart} agrees with the corresponding expression reported in Ref.~\cite{Cocco:2025adu} with the only difference that here $\hat{L}$ denotes the angular momentum divided by $M c$.

For our analysis, after averaging over the inner orbit, we also include in the Hamiltonian of the inner binary the $1$PN correction accounting for periastron precession~\cite{Kuntz:2021hhm,Kuntz_2021,Cocco:2025adu,Randall:2018qna}:
\begin{equation}
\label{H1PNave}
\left\langle H_{\rm 1PN}\right\rangle = -\mu\,\frac{3 G^2 M^2}{c^2 a^2 \sqrt{1 - e^2}}~,
\end{equation}
which is obtained by averaging the $1$PN Einstein–Infeld–Hoffmann Hamiltonian using Eqs.~\eqref{kuntzparametrization} and~\eqref{averageintegral}.
The complete expression for the averaged $1$PN Hamiltonian of an isolated binary can be found in Ref.~\cite{Damour:1988mr}.
Here, we neglect the term that depends only on the semi-major axis $a$, since it does not contribute to the LPE.
The contribution in Eq.~\eqref{H1PNave} can therefore be directly added to Eq.~\eqref{Hmarckaveinner}, as it represents a $1$PN correction to the internal dynamics of the binary, which is consistently treated only up to $0.5$PN order.
The $1$PN correction in Eq.~\eqref{H1PNave} induces a secular precession of the inner orbit, characterized by a timescale \cite{Kuntz:2021hhm}
\begin{equation}
    \label{eq:T1PN}
    T_{\rm 1PN} \sim u^{\hat t}\frac{a}{3}\left(\frac{a}{GM}\right)^{3/2}c^2 ,
\end{equation}
which sets the characteristic timescale for the inner periastron precession. Here, $u^{\hat{t}} := d\hat{t} / d\hat{\tau}$ is the gravitational redshift factor relating the proper time $\hat{\tau}$ along the outer geodesic to the coordinate time $\hat{t}$ of an asymptotic observer.
Consequently, the Hamiltonian of the system becomes:
\begin{equation}
    \label{Hmarckaveinner_plusprecession}
    \left\langle H_{\text{inner}}\right\rangle=-\frac{G M \mu}{2 a}+\left\langle H_\mathcal{E}\right\rangle+\left\langle H_\mathcal{B}\right\rangle+\left\langle H_{\rm 1PN}\right\rangle \,.
\end{equation}

At this stage, we briefly clarify the PN bookkeeping adopted in this work. The tidal interaction between the inner binary and the SMBH is treated at leading quadrupolar order, including the electric ($0$PN) and magnetic ($0.5$PN) tidal couplings that enter the precession-resonance condition.
To enable precession resonances and their crossing during the inspiral, we additionally retain the $1$PN contribution responsible for inner periastron precession and the $2.5$PN radiation-reaction (RR) terms (see Sec.~\ref{sec:Numerical}).
Although formally of higher PN order, these terms control the precession frequency and the resonance timing, respectively.
As discussed in Ref.~\cite{Kuntz:2021hhm}, we work in a regime that suppresses quadrupolar-driven secular effects (i.e., ZLK mechanism), so that higher-order multipolar and cross-terms are subleading on the resonance timescale and are consistently neglected.

\subsection{Validity of the inner-averaging approximation}
\label{subsec:validity_averaging}

The Hamiltonian of the inner binary derived in Eq.~\eqref{Hmarckaveinner_plusprecession} is averaged over the inner orbital motion. The validity of this inner-averaging approximation relies on a clear separation between the inner orbital period and all other timescales governing the evolution of the system.

The inner orbital timescale is defined as
\begin{equation}
\label{eq:Tinnorb}
T_{\rm in,orb} \sim u^{\hat t}\sqrt{\frac{a^3}{GM}},
\end{equation}
corresponding to the inverse Keplerian frequency of the inner binary, expressed in terms of the asymptotic time. This timescale must be much shorter than the characteristic timescales associated with the internal secular evolution of the binary, as well as the timescale over which the external tidal field varies along the outer orbit.

The secular evolution of the inner binary is governed by the relativistic periastron-precession timescale $T_{\rm 1PN}$, defined in Eq.~\eqref{eq:T1PN}, and by the RR timescale $T_{\rm in,RR}$ associated with gravitational-wave emission, which is discussed in detail in Sec.~\ref{sec:Numerical} (see Eq.~\eqref{PetersEQ}). For all configurations considered in this work, we verified that the inner orbital period remains much shorter than both of these secular timescales throughout the evolution. In particular, for the parameter sets shown in Figs.~\ref{fig1_compI}, \ref{fig1_aligned}, and~\ref{fig1_highhate}, we find $T_{\rm 1PN}/T_{\rm in,orb} \sim 10^2$ and $T_{\rm in,RR}/T_{\rm in,orb} \sim 10^7$.

% \Troels{I guess the above check is also required for the PN expansion to be valid? If $T_{\rm in,orb}$ would be similar to $T_{\rm 1PN}$ and/or $T_{\rm in,RR}$ it would presumably mean the break down of the PN expansion?}

% \martaC{Yes. Indeed, the timescale ratio depends on some power of the inner-semimajor axis, $a[t]$, and in our checks for the PN expansion to be valid we always made sure that $a\gg G M/c^2$}

In addition, the inner-averaging procedure requires that the inner orbital period should be much shorter than the shortest timescale over which the tidal field varies along the outer orbit. For nearly circular outer orbits this timescale is comparable with the outer orbital period, while for moderate-to-high outer eccentricities the dominant tidal forcing occurs during short pericenter passages. We therefore estimate the tidal-variation timescale at pericenter by considering scalar invariants constructed from the electric and magnetic tidal tensors, $\mathcal{E}_{ij}$ and $\mathcal{B}_{ij}$. Specifically, we define
\begin{equation}
T_{\mathcal{E},{\rm per}} \sim \sqrt{\frac{2\,\mathcal{E}_{ij}\mathcal{E}^{ij}}{\left|\partial_{\hat t}^2(\mathcal{E}_{ij}\mathcal{E}^{ij})\right|}},
\quad
T_{\mathcal{B},{\rm per}} \sim \sqrt{\frac{2\,\mathcal{B}_{ij}\mathcal{B}^{ij}}{\left|\partial_{\hat t}^2(\mathcal{B}_{ij}\mathcal{B}^{ij})\right|}},
\end{equation}
both evaluated at the outer pericenter passage. \footnote{At pericenter, the first time derivative of the scalar invariants $\mathcal{E}_{ij}\mathcal{E}^{ij}$ and $\mathcal{B}_{ij}\mathcal{B}^{ij}$ vanishes, being proportional to the radial velocity $\dot{\hat r}$, which is zero at this turning point of the outer motion. The second derivative therefore provides the leading, non-vanishing measure of the local timescale of variation.}
Throughout this work, we conservatively require a separation of at least one order of magnitude between the $T_{\rm in, orb}$ and this $T_{\rm \mathcal{E},per}, T_{\rm \mathcal{B},per}$.
For the parameter sets shown in Figs.~\ref{fig1_compI}, \ref{fig1_aligned}, and the left panel of Fig.~\ref{fig1_highhate}, we find that the inner orbital period is separated from the tidal-variation timescale by more than two orders of magnitude, with $T_{\mathcal{E},{\rm per}}/T_{\rm in,orb},\, T_{\mathcal{B},{\rm per}}/T_{\rm in,orb} \gtrsim 10^2$. The most conservative configuration corresponds to the right panel of Fig.~\ref{fig1_highhate}, for which $T_{\mathcal{E},{\rm per}}/T_{\rm in,orb}$ and $T_{\mathcal{B},{\rm per}}/T_{\rm in,orb}$ are of order $\mathcal{O}(10)$, still ensuring the validity of the inner-averaging approximation.

\section{Precession resonance in the strong-gravity regime: magnetic contributions}
\label{sec:analytic}

In this section, we analytically investigate precession resonances, focusing on the contribution of the magnetic tidal moments.
These resonances have previously been analyzed in the weak-field regime~\cite{Kuntz:2021hhm} and, more recently, in the strong-gravity regime for the interaction between the SMBH and the inner binary~\cite{Cocco:2025adu}, where only the quadrupolar electric tidal moments, entering at the $0$PN order, were included.
Here, we extend that analysis by incorporating the magnetic tidal coupling, which arises at the $0.5$PN order.
We begin by briefly reviewing the action–angle formalism and then perform an expansion in the outer orbit eccentricity $\hat{e}$, retaining terms up to first order in $\hat{e}$.

\subsection{Fundamental frequencies and the resonance condition}
\label{sec:aaM}

As the center of mass of the inner binary moves in the gravitational field of the non-rotating SMBH of mass $M_*$, its trajectory follows a geodesic in Schwarzschild spacetime.
Consequently, describing the outer motion requires solving the geodesic equations for the coordinates $\hat{t}(\hat{\tau})$, $\hat{r}(\hat{\tau})$, $\hat{\theta}(\hat{\tau})$, and $\hat{\phi}(\hat{\tau})$ as functions of the proper time $\hat{\tau}$ along the outer orbit.

Owing to the spherical symmetry of the background, the motion can be confined to the equatorial plane by setting $\hat{\theta} = \pi/2$, without loss of generality.
Since our goal is to provide an analytical description of a phenomenon that is potentially observable in GWs, we express all quantities in terms of coordinate time $\hat{t}$, which serves as the physical time parameter for distant observers.
The geodesic equation for $\hat{t}(\hat{\tau})$ is then used to re-express the evolution of the remaining variables as functions of $\hat{t}$.
The equations governing $\hat{r}(\hat{t})$ and $\hat{\phi}(\hat{t})$ depend on the conserved energy $\hat{E}$ and the angular momentum $\hat{L}$, expressed in units of $M c^2$ and $M c$, respectively.

In addition to the radial and azimuthal motion, an additional independent dynamical variable is Marck’s angle $\hat{\Psi}(\hat{t})$, which characterizes the evolution of a tetrad that is parallel transported along the geodesic.

Significant progress has been made in solving these equations analytically, particularly for bound orbits. Recent works have derived closed-form solutions in terms of elliptic functions~\cite{Fujita_2009, Schmidt_2002, van_de_Meent_2020, Drasco:2003ky, Hinderer_2008, Drasco_2006}.
Many of these results employ an {\it action–angle formalism}, in which the dynamics is expressed in terms of conserved actions $J_\mu$ and linearly evolving angles $q_i \equiv \Omega_i \hat{t}$, with $i = (\hat{r}, \hat{\phi}, \hat{\Psi})$. The rates $\Omega_i$ define the system’s {\it fundamental frequencies}.
In this framework, the solutions for the radial and azimuthal coordinates, as well as Marck’s angle, can be written in the following form~\cite{Fujita_2009, Schmidt_2002, van_de_Meent_2020, Drasco:2003ky}:
\begin{subequations}
        \begin{align}
        \label{eq:solr}
         \hat{r}(\hat{t}) =& f_{\hat r \hat r} \,, \\
         \label{eq:solphi}
         \hat{\phi}(\hat{t}) =& q_{\hat{\phi}} + f_{\hat \phi \hat r}  \,, \\ 
         \label{eq:solPsi}
        \hat{\Psi}(\hat{t}) =&  q_{\hat{\Psi}} + f_{\hat \Psi \hat r} \,,
         \end{align}
     \end{subequations}
where $f_{\hat{r}\hat{r}}$, $f_{\hat{\phi}\hat{r}}$, $f_{\hat{\Psi}\hat{r}}$ are elliptic functions of the radial angle variable $q_{\hat{r}}$.
     
The first terms in Eqs.~\eqref{eq:solphi} and~\eqref{eq:solPsi} describe the secular accumulation of $\hat{\phi}(\hat{t})$ and $\hat{\Psi}(\hat{t})$ over time, as expected for a purely circular orbit.
In contrast, the functions $f_{\hat{\phi} \hat{r}}$ and $f_{\hat{\Psi} \hat{r}}$, which are periodic in the radial motion, encode the oscillatory corrections arising from the orbit’s eccentricity.
The fundamental frequencies $\Omega_{\hat{r}}$, $\Omega_{\hat{\phi}}$, and $\Omega_{\hat{\Psi}}$ depend on the conserved quantities $\hat{E}$ and $\hat{L}$, but can equivalently be expressed in terms of the outer orbit’s semi-major axis $\hat{a}$ and eccentricity $\hat{e}$~\cite{Schmidt_2002}.

Since we are interested in the resonant dynamics of the triple system, we apply this formalism to rewrite the Hamiltonian in Eq.~\eqref{Hmarckaveinner} in terms of the fundamental frequencies of the outer orbit.
In this context, the only terms that depend on the outer dynamics are the quadrupolar tidal contributions, $\langle H_{\mathcal{E}}\rangle$ and $\langle H_{\mathcal{B}}\rangle$, defined in Eqs.~\eqref{eq:secEpart} and~\eqref{eq:secB}, respectively.
The analysis of the electric tidal Hamiltonian $\langle H_{\mathcal{E}}\rangle$ in terms of the fundamental frequencies was already carried out in Ref.~\cite{Cocco:2025adu}. Here, we focus on the magnetic contribution $\langle H_{\mathcal{B}}\rangle$, for which the nonzero components of $\mathcal{B}_{ij}$, as discussed in App.~\ref{Tidal_Moments_Schwarzschild}, are~\footnote{From now on, we drop the explicit $\hat{t}$-dependence of the variables $\hat{r}, \hat{\phi}, \hat{\Psi}$.}
\begin{equation}
    \begin{aligned} 
    \label{eq:B13_B23}
    &\mathcal{B}_{13}=\mathcal{B}_{31}=-\frac{3 \hat{r}_s \hat{L} \sqrt{\hat{L}^2+\hat{r}^2}}{2 \hat{r}^5}\cos{\hat \Psi}\,,\\
    &\mathcal{B}_{23}=\mathcal{B}_{32}=-\frac{3 \hat{r}_s \hat{L}\sqrt{\hat{L}^2+\hat{r}^2}}{2 \hat{r}^5}\sin{\hat \Psi} \,.
    \end{aligned}
\end{equation}     
Substituting the solutions for $\hat r$ and $\hat \Psi$ in terms of the angle variables $q_{\hat{r},\hat{\Psi}}$ given in Eqs.~\eqref{eq:solr} and \eqref{eq:solPsi} 
% in terms of the angle variables $q_{\hat{r},\hat{\Psi}}$ 
into Eq.~\eqref{eq:B13_B23}, 
and then performing a Fourier expansion, we get
\begin{equation}
    \label{eq:B13_B23_FOURIER}
    \begin{aligned} 
    &\mathcal{B}_{13}=\sum_{k, l}\left[m_{k,l} \cos (k q_{\hat{r}} + l q_{\hat{\Psi}})+\widetilde{m}_{k,l} \sin (k q_{\hat{r}} + l q_{\hat{\Psi}})\right]\,,\\
    &\mathcal{B}_{23}=\sum_{k, l}\left[n_{k,l} \sin (k q_{\hat{r}} + l q_{\hat{\Psi}})+\widetilde{n}_{k,l} \cos (k q_{\hat{r}} + l q_{\hat{\Psi}})\right]\,.
    \end{aligned}
\end{equation}

By projecting onto trigonometric functions of the resonant angles
\begin{equation}
\label{resonant angles}
\chi_{k,l} \equiv \gamma - (k\, \Omega_{\hat{r}} + l\, \Omega_{\hat{\Psi}})\hat{t}\,,
\end{equation}
which drive the precession resonances~\cite{Kuntz:2021hhm, Cocco:2025adu}, the magnetic Hamiltonian contributing to the resonant dynamics can be written as
\begin{equation} 
\label{eq:H_B_Fourier_generic}
    \left\langle H_\mathcal{B}^r\right\rangle=\frac{c}{3} \left\langle T_1\right\rangle\sum_{k, l}\left[ f_{k,l}(\vartheta,I)\cos{\chi_{k,l}}+g_{k,l}(\vartheta,I)\sin{\chi_{k,l}}\right]\,.
\end{equation}
Here, $H_\mathcal{B}^r$ denotes the resonant part of the magnetic Hamiltonian, $\langle T_1 \rangle$ is the quantity introduced in Eq.~\eqref{eq:averageTi}, and we define
\begin{equation}
    \label{eq:fkl_gkl_functions}
    \begin{aligned}
        f_{k,l}(\vartheta,I)&=\left( m_{k,l} \cos{I} + n_{k,l} \cos{2I} \right) \cos{\vartheta}\\
        &+\left(\tilde{n}_{k,l}\cos{I}-\tilde{m}_{k,l} \cos{2I} \right) \sin{\vartheta} \,, \\
        g_{k,l}(\vartheta,I)&= \left(-\tilde{m}_{k,l} \cos{I} +\tilde{n}_{k,l}\cos{2I}\right) \cos{\vartheta} \\&-\left(n_{k,l} \cos{I}+m_{k,l}\cos{2I}\right)\sin{\vartheta}\,.
    \end{aligned}
\end{equation}
We refer to $\chi_{k,l}$ as {\it resonant angles} because their vanishing gives rise to non-oscillatory terms in Eq.~\eqref{eq:H_B_Fourier_generic}, therefore they can drive a change in the inner orbital parameters.

This defines the resonance condition
\begin{equation}
\label{eq:res_condition_q1}
\dot{\gamma} = k\, \Omega_{\hat r} + l\, \Omega_{\hat \Psi}\,,
\end{equation}
which corresponds to Eq.~\eqref{eq:intro} for $q=1$.
This specific value of $q$ arises from the coupling between the magnetic tidal moments $\mathcal{B}_{ij}$ and the inner dynamical tensor $\mathcal{T}^{ij}$.
The dependence on $\gamma$ is encoded in $\mathcal{T}^{ij}$ through the components of $\boldsymbol{\hat u}$ and $\boldsymbol{\hat v}$, as defined in Eq.~\eqref{eq:uandv}.
From Eq.~\eqref{eq:Tij}, we see that $\mathcal{T}^{ij}$ involves only the rotation matrix $R_\gamma$ acting on $\boldsymbol{\hat u}$ and $\boldsymbol{\hat v}$, since the vector $(\boldsymbol{\hat u} \times \boldsymbol{\hat v})$ is independent of $\gamma$.
We note that in Ref.~\cite{Cocco:2025adu}, the resonance condition~\eqref{eq:intro} corresponds instead to $q=2$, consistent with the distinct $\gamma$-dependence of the tensor $\mathcal{Q}^{ij}$.

Since the angle $\hat{\Psi}$ enters the magnetic tidal moments in Eq.~\eqref{eq:B13_B23} only through $\cos{\hat{\Psi}}$ and $\sin{\hat{\Psi}}$, the index $l$ is restricted to the values $l = -1, 0, 1$.
Furthermore, because $\dot{\gamma} = d\gamma / d\hat{t}$ and both $\Omega_{\hat{r}}$ and $\Omega_{\hat{\Psi}}$ are positive, Eq.~\eqref{eq:res_condition_q1} constrains $k$ to satisfy $k\,\Omega_{\hat{r}} + l\,\Omega_{\hat{\Psi}} > 0$.
It is worth noting that the resonance condition is expressed solely in terms of the fundamental frequencies $\Omega_{\hat{r}}$ and $\Omega_{\hat{\Psi}}$.
This remains the case even when precession resonances are analyzed in a different reference frame, such as the distant-star frame introduced in App.~\ref{App:DistantStarFrame}, as already emphasized in Ref.~\cite{Cocco:2025adu}.

\subsection{Perturbative approach for quasi-circular outer orbits} \label{PerturbativeApproach}

We have shown that the description of the outer motion using action-angle variables provides a powerful framework for identifying the multiperiodic structure of geodesic motion in curved spacetime. In our setup, this formalism applies to the motion of the inner binary's CoM around a Schwarzschild SMBH. This enables us to express the magnetic quadrupole contribution to the Hamiltonian in Eq.~\eqref{eq:secB}, in terms of the corresponding fundamental frequencies, as given in Eqs.~\eqref{eq:B13_B23_FOURIER}-\eqref{eq:fkl_gkl_functions}, and to derive the resonance condition in Eq.~\eqref{eq:res_condition_q1}. 
The resulting structure exhibits a richer spectrum of resonances than the analogous relativistic condition found in Ref.~\cite{Cocco:2025adu} for the case of electric tides.

However, the analytical solutions for $\hat{r}$ and $\hat{\Psi}$ given in Eqs.~\eqref{eq:solr} and~\eqref{eq:solPsi} involve elliptic functions, which complicate both the explicit computation of the Fourier coefficients in Eq.~\eqref{eq:B13_B23_FOURIER} and the identification of the $(k,l)$ combinations that effectively contribute to the resonance condition~\eqref{eq:res_condition_q1}.
The former determines the amplitude of each resonance peak, while the latter specifies the point in the system’s evolution, namely the value of the semi-major axis $a$, at which the inner binary crosses each precession resonance.
Both ingredients are therefore essential for developing a predictive analytical model that avoids reliance on full numerical integration.

To make analytical progress, we follow Ref.~\cite{Cocco:2025adu} and perform a quasi-circular expansion in the outer eccentricity, $\hat{e} \ll 1$, keeping the outer semi-major axis $\hat{a}$ and the angle variables $q_{\hat{r}, \hat{\Psi}}$ fixed. At leading order in this expansion, the solutions in Eqs.~\eqref{eq:solr} and~\eqref{eq:solPsi} reduce to~\cite{Cocco:2025adu}
\begin{equation}
\label{eq:randPsismalle}
    \begin{split}
        \hat{r}&=\hat a\left(1-\hat e \cos q_{\hat{r}}\right)+\mathcal{O}\left(\hat e^2\right)\,,
        \\
        {\hat \Psi}&=q_{{\hat\Psi}}+2 \hat e \frac{\hat{\sigma}-4}{\hat{\sigma}-2} \sqrt{\frac{\hat{\sigma}-3}{\hat{\sigma}-6}} \sin q_{\hat{r}}+\mathcal{O}\left(\hat e^2\right)\,,
    \end{split}
\end{equation}
 where 
    \begin{equation}
        \hat{\sigma} = \frac{2 \hat{a}}{\hat{r}_s}
    \end{equation}
    is a dimensionless parameter measuring relativistic corrections: it is finite in the strong-field regime, while $\hat{\sigma}\rightarrow \infty$ in the weak-field limit. We recall that the angle variables are defined as $ q_{\hat r, \hat{\Psi}} \equiv \Omega_{\hat{r}, \hat{\Psi}}\,\hat{t}$, with the corresponding frequencies given by~\cite{Cocco:2025adu} 
\begin{subequations}
    \begin{align}
    \label{eq:omegarsmalle}
    \Omega_{\hat{r}}&= \Omega_{\rm 0PN} \sqrt{\frac{\hat{\sigma}-6}{\hat{\sigma}}}+\mathcal{O}(\hat{e}^2) \,,
    \\
    \label{eq:omegaPsismalle}
    \Omega_{\hat{\Psi}} & = \Omega_{\rm 0PN} \sqrt{\frac{\hat{\sigma}-3}{\hat{\sigma}}} +\mathcal{O}(\hat{e}^2) \,,
    \end{align}
\end{subequations}
where $\Omega_{\rm 0PN} \equiv \sqrt{G M_* / \hat{a}^3}$ is the Keplerian frequency of the outer motion. As expected, both relativistic frequencies in Eqs.~\eqref{eq:omegarsmalle},~\eqref{eq:omegaPsismalle} reduce to $\Omega_{\rm 0PN}$ in the weak-field limit $\hat{\sigma}\rightarrow \infty$.

By substituting Eq.~\eqref{eq:randPsismalle} into Eq.~\eqref{eq:B13_B23}, we find that, to leading order in $\hat{e}$, the only non-vanishing Fourier coefficients in the expansion~\eqref{eq:H_B_Fourier_generic} are
\begin{subequations}
    \begin{align}
    \label{eq:fouriercoeffsmalle-11}
        & m_{-1,1}=n_{-1,1}= \notag \\
        &-\frac{\hat{e}}{\hat{r}_s^2 \hat{\sigma}^3}\frac{6\left(4\hat{\sigma}-7-2 \sqrt{\frac{\hat{\sigma}-3}{\hat{\sigma}-6}}(\hat{\sigma}-4)\right)}{(\hat{\sigma}-3) \sqrt{\hat{\sigma}-2}} + \mathcal{O}(\hat{e}^2)\,, \\
    \label{eq:fouriercoeffsmalle01}
        & m_{0,1}=n_{0,1}=-\frac{1}{\hat{r}_s^2 {\hat \sigma}^3}\frac{12 \sqrt{\hat{\sigma}-2}}{\hat{\sigma}-3} + \mathcal{O}(\hat{e}^2)\,, \\
    \label{eq:fouriercoeffsmalle11}
        & m_{1,1}=n_{1,1}= \notag\\
        &-\frac{\hat{e}}{\hat{r}_s^2 \hat{\sigma}^3}\frac{6\left(4\hat{\sigma}-7+2 \sqrt{\frac{\hat{\sigma}-3}{\hat{\sigma}-6}}(\hat{\sigma}-4)\right)}{(\hat{\sigma}-3) \sqrt{\hat{\sigma}-2}} + \mathcal{O}(\hat{e}^2)~~.
    \end{align}
\end{subequations}
We note that the coefficients $\tilde{m}_{k,l}$ and $\tilde{n}_{k,l}$ in Eqs.~\eqref{eq:B13_B23_FOURIER} vanish due to the parity properties of the expansion of $\mathcal{B}_{ij}$ at linear order in $\hat{e}$.
Moreover, the coefficients in Eqs.~\eqref{eq:fouriercoeffsmalle-11}–\eqref{eq:fouriercoeffsmalle11} vanish in the weak-field limit $\hat{\sigma}\rightarrow\infty$, a behavior which is consistent with the purely relativistic nature of the magnetic tidal moments. These coefficients indicate that, in the quasi-circular limit of the outer orbit, the resonance condition~\eqref{eq:res_condition_q1} reduces to
\begin{equation}
    \begin{aligned} \label{magneticresonancecondition}
        & \dot{\gamma}=-\Omega_{\hat{r}}+\Omega_{\hat\Psi}\,,\\
        & \dot{\gamma}=\Omega_{\hat\Psi}\,, \\
        & \dot{\gamma}=\Omega_{\hat{r}}+\Omega_{\hat\Psi}\,,
    \end{aligned}
\end{equation}
each corresponding to a distinct peak in the evolution of the inner eccentricity.
Consequently, retaining terms up to first order in $\hat{e}$, Eq.~\eqref{eq:H_B_Fourier_generic} simplifies to
\begin{equation} 
\label{eq:H_B_Fourier_FirstOrder}
    \begin{aligned}
        \left\langle H_\mathcal{B}^r\right\rangle &=   \mu c \frac{m_1-m_2}{M}e a\sqrt{G M a (1-e^2)} \cos^2{\frac{I}{2}}(1-2\cos{I}) \\ & \times \sum_{k,\,l} m_{k,\,l}(\cos{\vartheta\cos{\chi_{k,\,l}}}
        -\sin{\vartheta}\sin{\chi_{k,\,l}})\,,
    \end{aligned}
\end{equation}
where $(k,l)$ are now restricted to the values $(-1,1)$, $(0,1)$ and $(1,1)$.

The Hamiltonian in Eq.~\eqref{eq:H_B_Fourier_FirstOrder} reveals how precession resonances depend on the relative orientation of the inner binary with respect to the outer SMBH of mass $M_*$.
Specifically, for small outer eccentricities $\hat e$, no resonances occur when $I = \pi$, corresponding to an anti-aligned configuration of the inner and outer angular momenta.
Interestingly, resonances also vanish at $I = \pi/3$.
In contrast, when the angular momenta are aligned ($I = 0$), the resonant effects arising from the $0.5$PN magnetic coupling in Eq.~\eqref{eq:H_B_Fourier_FirstOrder} are not suppressed.
In fact, they are enhanced, since $I = 0$ corresponds to a global maximum of $\langle H_{\mathcal{B}}^r \rangle$ for fixed $\vartheta$.
This behavior is particularly noteworthy, as certain precession resonances driven by the $0$PN electric tidal coupling are instead suppressed at zero inclination~\cite{Cocco:2025adu}.

It is important to emphasize that all results from Eq.~\eqref{eq:randPsismalle} to Eq.~\eqref{eq:H_B_Fourier_FirstOrder}, including their astrophysical implications, have been obtained within the small–outer-eccentricity approximation. Consequently, they are expected to accurately describe systems where the outer orbit remains nearly circular, such as b-EMRIs in the late stages of orbital circularization.

% \section{Numerical solution to the LPE equations}

\section{Numerical solution to the evolution equations of the triple system}

\label{sec:Numerical}

In this section, we present numerical solutions for the evolution of the inner binary’s orbital parameters.
By solving the full equations of motion, without performing a Fourier expansion of the Hamiltonian, we verify the analytical results derived in Sec.~\ref{PerturbativeApproach} for small values of the outer eccentricity $\hat e$.
More generally, our numerical framework allows us to explore configurations with finite $\hat e$.
However, the present numerical analysis is best suited to astrophysical systems with modest $\hat e$, as high-eccentricity formation channels may require an impulsive or periastron-kick description which is beyond the scope of this work.
To isolate the contribution of precession resonances, we adopt a hierarchy of timescales $T_{\rm 1PN} \ll T_{\rm ZLK}$, which suppresses the standard ZLK oscillations even for large initial inclinations between the inner and outer orbits~\cite{Kuntz:2021hhm,Camilloni_2024,Naoz_2016,Cocco:2025adu}.

\subsection{Inner binary}

The evolution of the inner binary is governed by the LPE, supplemented by the gravitational radiation–reaction (RR) terms at the leading $2.5$PN order described by Peters’ equations~\cite{Peters, PetersMathews}
\begin{equation}
\begin{split} 
    \label{PetersEQ}
    \frac{d a}{d \hat{t}} \Bigg|_{\mathrm{GW}}& = -\frac{1}{u^{\hat{t}}}\frac{64}{5} \frac{G^3 \mu~ M^2}{c^5 a^3} \frac{1}{\left(1-e^2\right)^{7 / 2}} \left(1+\frac{73}{24} e^2+\frac{37}{96} e^4\right), \\
    \frac{d e}{d \hat{t}} \Bigg|_{\mathrm{GW}}& = -\frac{1}{u^{\hat{t}}}\frac{304}{15} \frac{G^3 \mu ~ M^2}{c^5 a^4} \frac{e}{\left(1-e^2\right)^{5 / 2}}\left(1+\frac{121}{304} e^2\right) .
\end{split}
\end{equation}
The gravitational back-reaction causes the inner binary to lose energy and angular momentum through gravitational-wave emission, leading to a gradual reduction of its semi-major axis $a$ and eccentricity $e$.
As a result, the inner orbit shrinks and circularizes as the system evolves toward merger, which in turn accelerates the periastron precession and drives the binary to naturally cross multiple precession resonances during its inspiral.

This process occurs on the so-called RR timescale, $T_{\rm in,RR}$, which characterizes the secular orbital decay induced by GW emission. For a binary with initial semi-major axis $a_0$ and small eccentricity $e_0$, $T_{\rm in,RR}$ can be estimated from Peters’ equations~\eqref{PetersEQ} and approximately reads~\cite{maggiorevol1}
\begin{equation} 
    \label{T_RR_approx}
    T_{\rm in,RR}= u^{\hat{t}}\frac{5 a_0^4 c^5}{256 G^3 m_1 m_2 M} \,.
\end{equation}
Note that the $2.5$PN gravitational back-reaction is the leading dissipative effect, and since we model the inner binary up to $0.5$PN order, we can simply add this contribution to the set of evolution equations for the inner binary's orbital parameters. It is worth noting that environmental effects can introduce dissipative contributions to the evolution of the inner binary, thereby interacting with the RR process \cite{Cardoso:2020iji, Zwick:2021ayf, CanevaSantoro:2023aol,Ishibashi:2024wwu,Barausse:2007dy}. Here, we neglect these effects, but in App.~\ref{App:Environmental effects} we briefly explore this scenario.

We recall from Sec.~\ref{sec:orbital_gauge} that we adopt the generalized Lagrange gauge, in which the LPE for the Delaunay variables form a symplectic system~\cite{gaugesymmetry,EfroimskyGoldreichHJapproach}.
From this formulation, the LPE expressed in terms of the Keplerian elements can be obtained by rewriting $a$, $e$, and $I$ in terms of the Delaunay action variables defined in Eq.~\eqref{actionvariables} (see~\cite{Valtonen_Karttunen_2006} for details).
The resulting full set of equations governing the evolution of the inner binary is therefore given by
\begin{equation}
\begin{aligned} 
\label{LPEpeters}
\frac{d a}{d \hat{t}} & = \frac{1}{u^{\hat{t}}}\left(-\sqrt{\frac{4 a}{G M}} \frac{\partial \mathcal{H}}{\partial \beta}\right.
\\&\left. \phantom{}-\frac{64}{5} \frac{G^3 \mu~ M^2}{c^5 a^3} \frac{1}{\left(1-e^2\right)^{7 / 2}}
\left(1+\frac{73}{24} e^2+\frac{37}{96} e^4\right)\right) \,, \\
\frac{d e}{d \hat{t}} & =\frac{1}{u^{\hat{t}}}\left(\sqrt{\frac{1-e^2}{G M a e^2}} \frac{\partial \mathcal{H}}{\partial \gamma}-\frac{1-e^2}{\sqrt{G M a}e} \frac{\partial \mathcal{H}}{\partial \beta}\right.\\
&\left.\phantom{}-\frac{304}{15} \frac{G^3 \mu ~ M^2}{c^5 a^4} \frac{e}{\left(1-e^2\right)^{5 / 2}}\left(1+\frac{121}{304} e^2\right)\right) \,, \\
\frac{d I}{d \hat{t}} & =\frac{1}{u^{\hat{t}}}\frac{1}{\sqrt{G M a\left(1-e^2\right)} \sin I} \left( \frac{\partial \mathcal{H}}{\partial \vartheta}- \cos I \frac{\partial \mathcal{H}}{\partial \gamma}\right) \,, \\
\frac{d \beta}{d \hat{t}} & =\frac{1}{u^{\hat{t}}}\left(\sqrt{\frac{4 a}{G M}} \frac{\partial \mathcal{H}}{\partial a}+\frac{1-e^2}{\sqrt{G M a} e} \frac{\partial \mathcal{H}}{\partial e}\right) \,, \\
\frac{d \gamma}{d  \hat{t}} & =\frac{1}{u^{\hat{t}}}\left(-\sqrt{\frac{1-e^2}{G M a e^2}} \frac{\partial \mathcal{H}}{\partial e}+\frac{\cot I}{\sqrt{G M a\left(1-e^2\right)}} \frac{\partial \mathcal{H}}{\partial I}\right) \,,\\
\frac{d \vartheta}{d \hat{t}} & =\frac{1}{u^{\hat{t}}}\left(-\frac{1}{\sqrt{G M a\left(1-e^2\right)} \sin I} \frac{\partial \mathcal{H}}{\partial I}\right) \,.
\end{aligned}
\end{equation}
Here, $\mathcal{H}$ denotes the complete averaged Hamiltonian of the inner binary, given in Eq.~\eqref{Hmarckaveinner_plusprecession} and rescaled by $\mu$, while we recall that $u^{\hat{t}} = d\hat{t} / d\hat{\tau}$ is the gravitational redshift factor.

\subsection{Outer binary}

To fully characterize the evolution of the inner binary, we must also account for its motion along the outer orbit around the SMBH.
To this end, we parameterize the radial coordinate $\hat{r}$ of the outer orbit, in analogy with Newtonian dynamics, in terms of its semi-major axis $\hat{a}$ and eccentricity $\hat{e}$~\cite{Chandrasekhar:1985kt}
\begin{equation}
\label{rhatparametr}
    \hat{r} = \frac{\hat{a}(1-\hat{e}^2)}{1+\hat{e}\cos \hat{\psi}} \,.
\end{equation}
Here, $\hat{\psi}$ denotes the relativistic anomaly, whose evolution with respect to the asymptotic time $\hat{t}$ is given by~\cite{Chandrasekhar:1985kt}
\begin{equation}
    \label{DpsiDt}
    \begin{split}
    \frac{d \hat{\psi}}{d \hat{t}}= & \sqrt{\frac{G M_*}{\hat{a}^{3}(1-\hat{e}^2)^{3}}} \frac{ \left(1+\hat{e} \cos \hat{\psi}\right)^{2}}{\sqrt{(2 \delta -1)^2- 4 \delta^2 \hat{e}^2}} \\ 
    & \times \sqrt{1-2 \delta (3+ \hat{e} \cos \hat{\psi})} \left(1-2\delta (1+\hat{e} \cos\hat{\psi})\right) \,,
    \end{split}
\end{equation}
where $\delta= G M_* / (c^2 \hat{a} (1-\hat{e}^2) )$. Eqs.~\eqref{rhatparametr} and~\eqref{DpsiDt} together describe the evolution of the radial coordinate $\hat{r}$ as the inner binary moves along its outer orbit.
Since we are working in the local inertial frame, we must also account for the evolution of Marck’s angle $\hat{\Psi}$, which tracks the orientation of the tetrad defining our reference frame.
Its evolution with respect to the asymptotic time $\hat{t}$ is given by~\cite{Marck:1983}\footnote{We remind the reader that, in our notation, $\hat{L}$ has dimensions of length.}
\begin{equation}
    \label{DPsiDt}
     \frac{d{\hat\Psi}}{d\hat{t}}= \frac{c}{u^{\hat{t}}}\frac{\hat{E}\hat{L}}{\hat{r}^2+\hat{L}^2} \, .
\end{equation}

As anticipated in Sec.~\ref{sec:1/2PNsystem}, throughout this work we neglect the GW emission from the outer orbit and therefore treat the outer orbital parameters $\hat E$ and $\hat L$ (or equivalently $\hat a$ and $\hat e$) as constants during the inspiral of the inner binary. Here, we provide an explicit timescale estimate to assess the validity of this approximation.

The RR timescale associated with the inner binary, $T_{\rm in,RR}$, is set by GW emission from the inner orbit and is described by the Peters equations (see Eq.~\eqref{T_RR_approx}). As discussed above, resonant excitations can increase the inner eccentricity and shorten the effective inspiral time, so that $T_{\rm in,RR}$ should be regarded as a conservative upper bound.

The outer binary $(M_*,M)$ constitutes an extreme mass-ratio inspiral (EMRI). On timescales much shorter than RR, the motion of the inner binary’s center of mass is well approximated by a bound Schwarzschild geodesic, while GW emission drives an adiabatic evolution of the outer constants of motion on longer timescales. We estimate the outer RR timescale as \cite{Hinderer_2008}
\begin{equation}
\label{eq:ToutRR}
T_{\rm out,RR} \sim \left(\frac{M}{M_*}\right)^{-1} T_{\rm out,orb},
\end{equation}
where the characteristic outer orbital period is defined in terms of the azimuthal orbital frequency $\Omega_{\hat\phi}$ of the geodesic,
\begin{equation}
T_{\rm out,orb} \sim \frac{2\pi}{\Omega_{\hat\phi}}.
\end{equation}
This choice is physically motivated, since GW energy and angular-momentum fluxes are governed by the actual orbital motion of the center of mass around the SMBH. By contrast, the Marck frequency $\Omega_{\hat\Psi}$ controls the rotation of the parallel-transported tetrad and determines the time dependence of the tidal moments entering the resonance conditions.

For the fiducial systems shown in Figs.~\ref{fig1_compI},~\ref{fig1_aligned}, and the left panel of Fig.~\ref{fig1_highhate}, we find that the outer RR timescale typically exceeds the inner one by a factor of a few, with $T_{\rm out,RR}/T_{\rm in,RR} \sim 3 \,-\, 5$, while becoming comparable only in the most compact configuration shown in the right panel of Fig.~\ref{fig1_highhate}. Since resonant effects are concentrated in a fraction of the total inspiral, comparing $T_{\rm out,RR}$ to the time spent in the resonance-rich region improves this separation by a modest factor.

%\sout{We therefore conclude that including the RR of the outer orbit could quantitatively shift the timing of resonance crossings, particularly for systems near the ISO. While this effect is beyond the scope of the present work, which aims to isolate and characterize the role of precession resonances in the inner binary, it represents a natural refinement of our analysis that we leave for future work.}
We therefore conclude that including the RR of the outer orbit could quantitatively shift the timing of resonance crossings, particularly for systems near the ISO, where the hierarchy between the inner and outer RR timescales is only moderate.
Since resonance-crossing locations are sensitive to the evolution of the outer frequencies, a self-consistent treatment of the coupled inner-outer inspiral will be needed for future waveform-template applications.
This refinement is beyond the scope of the present work, which aims to isolate and characterize the role of precession resonances in the inner binary.

The evolution equations~\eqref{LPEpeters} for the orbital parameters of the inner binary, together with the evolution equations~\eqref{rhatparametr}, \eqref{DpsiDt} and \eqref{DPsiDt} for the orbital parameters of the outer orbit, form a complete set of coupled ordinary differential equations in terms of the asymptotic time $\hat{t}$. 
These equations can be integrated numerically, given suitable initial conditions, to follow the evolution of the inner binary as it orbits the SMBH.

Given the nature of the system, a compact binary in a tight orbit around a SMBH, it is essential to verify that the inner binary remains stable against tidal disruption throughout its evolution.
To ensure this, we impose the condition that the tidal forces exerted by the SMBH remain weaker than the internal binding forces of the inner binary, following the criterion introduced in Ref.~\cite{Suzuki:2020vfw}
\begin{equation}
\label{eq:stability}
     a \lesssim \frac{1}{4} \hat a \left(1-\hat{e}\right)\left(\frac{M}{3 M_*}\right)^{1/3} \,.
\end{equation} 
This stability criterion is similar to those adopted in Refs.~\cite{Antonini:2012ad, Miller_2005}, but follows the formulation of Ref.~\cite{Suzuki:2020vfw}, which incorporates the relativistic description of the SMBH and provides a conservative stability threshold.

Moreover, we compute the numerical solutions for configurations in which the outer motion corresponds to a stable bound orbit.
Specifically, we choose the semi-major axis $\hat{a}$ to be larger than that of the innermost stable orbit (ISO), namely~\cite{Brik_Geyer_Hinderer,maggiorevol1}
\begin{equation} 
    \label{eq:a_iso}
    \hat{a}_{\mathrm{ISO}}=\left(\frac{6+2 \hat{e}}{1-\hat{e}^2}\right) \frac{\hat{r}_s}{2} \,.
\end{equation}

\subsection{Results and physical interpretation}

The goal of our analysis is to investigate the role of the magnetic tidal moments, corresponding to the $0.5$PN terms in the Hamiltonian~\eqref{Hmarckaveinner_plusprecession}, in driving precession resonances.
To this end, Table~\ref{tab:gamma_frequencies} summarizes the resonance conditions associated with the electric tidal moments at $0$PN order, as derived analytically in Ref.~\cite{Cocco:2025adu}, and those produced by the magnetic tidal moments at $0.5$PN order, obtained analytically in Sec.~\ref{PerturbativeApproach}.
Both sets of conditions are evaluated at leading order in the outer-orbit eccentricity $\hat{e}$, and define the resonance condition
\begin{equation} \label{eq:mergedresonancecondition}
    \dot{\gamma} = k\,\Omega_{\hat{r}} + l\,\Omega_{\hat{\Psi}}~, \quad k\in \mathbb{Z}/2,\; l\in\mathbb{Z} \,.
\end{equation}

\begin{table}[ht]
    \centering
    \caption{Resonances condition at the first order in $\hat{e}$}
\begin{tabular*}{\linewidth}{
  @{\extracolsep{\fill}}
  c  
  c 
  c  
  c 
  @{}
}
\toprule
\multicolumn{4}{c}{
  $\dot{\gamma} = k\,\Omega_{\hat{r}} + l\,\Omega_{\hat{\Psi}},\;
  k\in \mathbb{Z}/2,\;
  l\in\mathbb{Z}$
} \\
\midrule
& $\mathcal{E}_{ij}\,(0\mathrm{PN})$
& $\mathcal{B}_{ij}\,(0.5\mathrm{PN})$ & \\
\midrule
& $\dot{\gamma} = \tfrac12\,\Omega_{\hat{r}}$ & & \\[1ex]
& $\dot{\gamma} = -\tfrac12\,\Omega_{\hat{r}} + \Omega_{\hat{\Psi}}$
& $\dot{\gamma} = -\Omega_{\hat{r}} + \Omega_{\hat{\Psi}}$ & \\[1ex]
& $\dot{\gamma} = \Omega_{\hat{\Psi}}$
& $\dot{\gamma} = \Omega_{\hat{\Psi}}$ & \\[1ex]
& $\dot{\gamma} = \tfrac12\,\Omega_{\hat{r}} + \Omega_{\hat{\Psi}}$
& $\dot{\gamma} = \Omega_{\hat{r}} + \Omega_{\hat{\Psi}}$ & \\
    \bottomrule
\end{tabular*}
    \label{tab:gamma_frequencies}
\end{table}

 Moreover, since the $0.5$PN term in the Hamiltonian~\eqref{HBmarckaveinner} is proportional to the mass difference between the two components of the inner binary, our numerical analysis focuses on asymmetric systems, i.e., $m_1 \neq m_2$.
This contrasts with the purely electric tidal interaction studied in Ref.~\cite{Cocco:2025adu}, where the term $\langle H_\mathcal{E} \rangle$ is independent of the mass difference, allowing the equal-mass case to be consistently considered.

More specifically, we compare the evolution of the inner binary’s eccentricity in two cases: one including only the $0$PN tidal coupling (black curve), and the other including both the $0$PN and $0.5$PN tidal couplings (blue curve). 

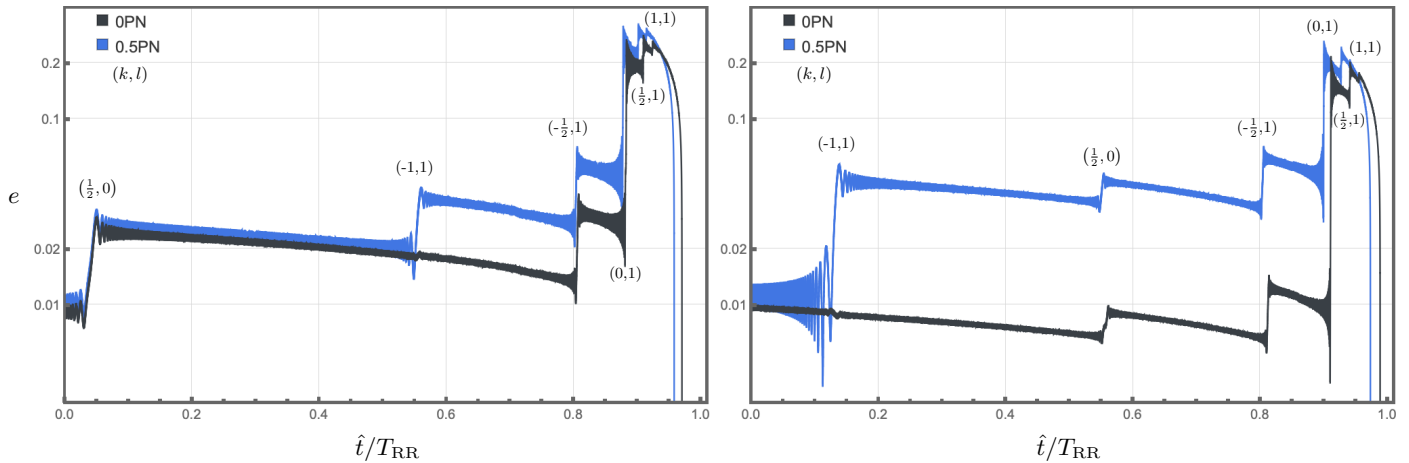
\begin{figure*}[ht]
    \centering
    \includegraphics[width=1\textwidth]{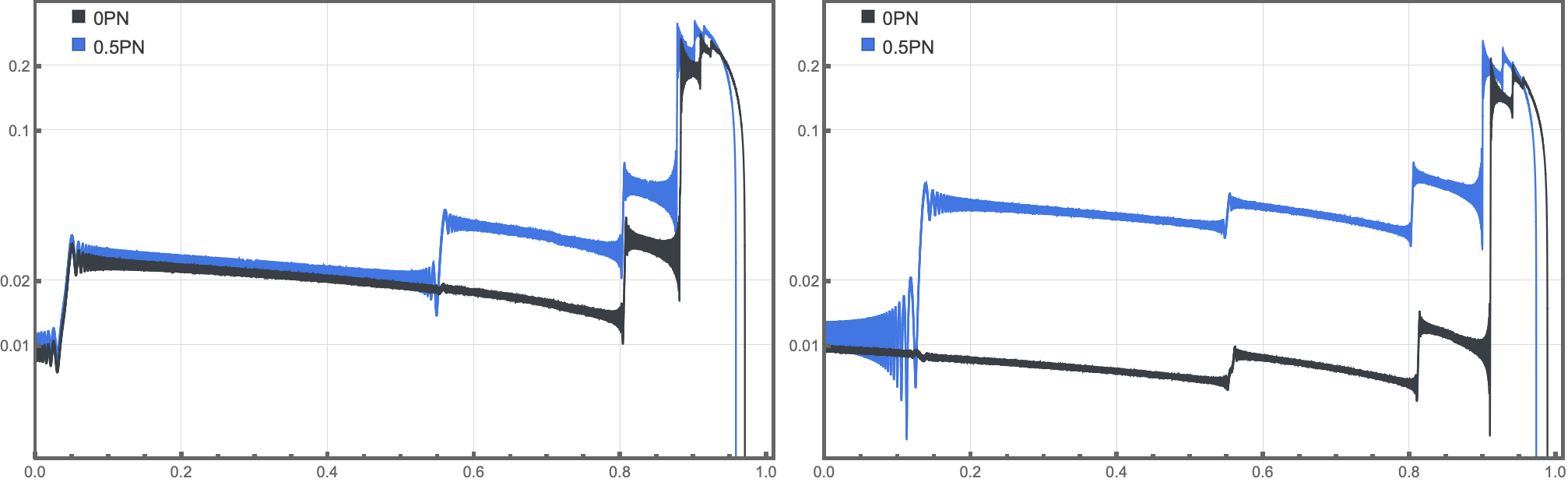}
    \begin{picture}(0,0)
        \put(-135,-2){\small $\hat t/T_{\rm in,RR}$  }
        \put(120,-2){\small $\hat t/T_{\rm in,RR}$  }
        \put(-265,93){$e$}
        \put(-226,140){\scalebox{0.7}{$(k,l)$}}
        \put(-239,97){\scalebox{0.65}{$\left(\frac{1}{2},0\right)$}}
        \put(-63,120){\scalebox{0.65}{(-$\frac{1}{2}$,1)}}
        \put(-40,65){\scalebox{0.65}{(0,1)}}
        \put(-32,132){\scalebox{0.65}{($\frac{1}{2}$,1)}}
        \put(-120,105){\scalebox{0.65}{(-1,1)}} 
        \put(-27,161){\scalebox{0.65}{(1,1)}}

        \put(30,140){\scalebox{0.7}{$(k,l)$}}
        \put(137,109){\scalebox{0.65}{$\left(\frac{1}{2},0\right)$}}
        \put(195,120){\scalebox{0.65}{(-$\frac{1}{2}$,1)}}
        \put(221,158){\scalebox{0.65}{(0,1)}}
        \put(231,123){\scalebox{0.60}{($\frac{1}{2}$,1)}}
        \put(38,114){\scalebox{0.65}{(-1,1)}} 
        \put(237,150){\scalebox{0.65}{(1,1)}}
    \end{picture}
   \caption{Evolution of the eccentricity $e$ of the inner binary as a function of the normalized time $\hat{t}/T_{\rm in,RR}$,
comparing the case with only the $0$PN tidal coupling (black curve) to that including both the $0$PN and $0.5$PN tidal couplings (blue curve), for a quasi-circular outer orbit.
The left panel corresponds to a region closer to the ISO of the SMBH, while the right panel shows a region farther from it.
In the former, the first precession resonance encountered by the inner binary arises solely from the $0$PN term in the Hamiltonian~\eqref{Hmarckaveinner_plusprecession}; in the latter, the first resonance is due exclusively to the magnetic $0.5$PN contribution.
The parameters and initial conditions are: $M_*=4\times 10^6\, \rm M_{\odot}$, $m_1=25\rm M_{\odot}$, $m_2=1.5\, \rm M_{\odot}$, $\hat e=0.08$, $\hat a=7.5 GM_*/c^2 \sim 0.3\,\text{AU}$ (left panel), $\hat a=10 G M_*/c^2 \sim 0.4 \text{AU}$ (right panel), $T_{\rm in,RR}\sim 6\,\text{yrs}$, GW peak frequency $f_{\mathrm{GW}}\sim0.06\, \text{Hz}$, $a_0\sim 0.0003\, \text{AU}$ (close to the first resonance), $e_0=0.01$, $\gamma_0=0^{\circ}$, $I_0=20^{\circ}$, $\vartheta_0=0^{\circ}$, $\hat\Psi_0=0^{\circ}$.
Numbers in parentheses indicate the $(k,l)$ pairs corresponding to each resonance peak, which satisfy the resonance condition~\eqref{eq:res_condition_q1}.
} 
    \label{fig1_compI}
\end{figure*}

In Fig.~\ref{fig1_compI}, we show the evolution of the inner binary’s eccentricity for a quasi-circular outer orbit, i.e., for small values of the outer eccentricity $\hat{e}$.
This configuration enables us to test the validity of the analytical model for precession resonances developed in Sec.~\ref{PerturbativeApproach}, by reproducing all the resonances $\left(k,l\right)$ listed in Table~\ref{tab:gamma_frequencies}.
We find that for small values of $\hat{e}$, the first precession resonance encountered by the inner binary depends on the semi-major axis $\hat{a}$ of the outer orbit, and can originate from either the $0$PN electric tidal moments or the $0.5$PN magnetic tidal moments. Specifically, when $\hat{a}$ is close to the ISO of the SMBH, the first precession resonance corresponds to the $\left(\tfrac{1}{2},0\right)$ resonance, driven by the electric tidal moments (left panel of Fig.~\ref{fig1_compI}).
Conversely, for larger values of $\hat{a}$, farther from the ISO, the first resonance encountered by the inner binary is the $\left(-1,1\right)$ resonance, which arises from the magnetic tidal moments (right panel of Fig.~\ref{fig1_compI}).\footnote{Even in this case, the system remains very close to the ISO of the SMBH, where the inner binary dynamics is strongly influenced by relativistic effects.}

\begin{figure*}[ht]
    \centering
    \includegraphics[width=1\textwidth]{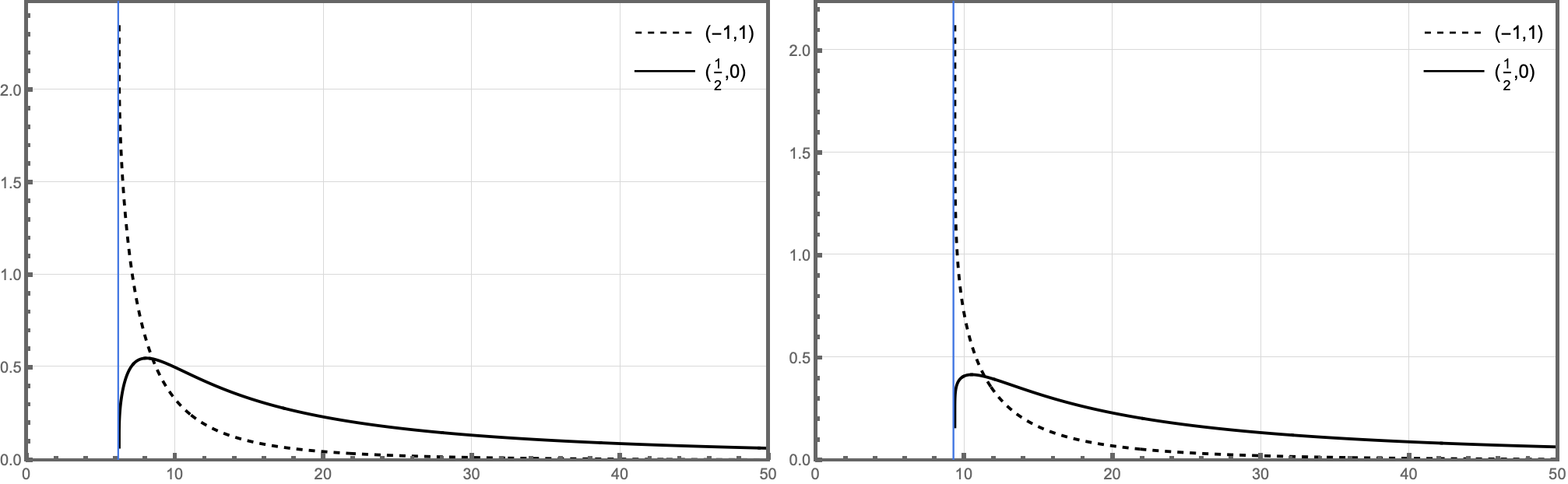}
    \begin{picture}(0,0)
        \put(-130,-2){\small $\hat a$  }
        \put(130,-2){\small $\hat a$  }
        \put(-270,80){\rotatebox{90}{mHz}}
    \end{picture}
    \caption{
Evolution of the first precession resonance encountered by the inner binary as a function of the outer semi-major axis $\hat{a}$, expressed in units of $GM_*/c^2$.
The solid black curve corresponds to the $\dot{\gamma} = \tfrac{1}{2}\Omega_{\hat{r}}$ resonance, i.e., $(k, l) = \left(\tfrac{1}{2}, 0\right)$, which arises solely from the $0$PN electric coupling in the Hamiltonian~\eqref{Hmarckaveinner_plusprecession}.
The dashed black curve represents the $\dot{\gamma} = -\Omega_{\hat{r}} + \Omega_{\hat{\Psi}}$ resonance, i.e., $(k, l) = (-1, 1)$, which originates from the $0.5$PN magnetic coupling.
In the left panel, we set $\hat{e} = 0.08$, while in the right panel $\hat{e} = 0.5$.
In both panels, the blue solid line marks the semi-major axis of the ISO, defined in Eq.~\eqref{eq:a_iso}, for a SMBH of mass $M_* = 4 \times 10^6\,M_\odot$. Notice that, for a fixed outer semi-major axis $\hat{a}$, the inner binary evolves upward along the y-axis.
}
\label{fig_freq}
\end{figure*}
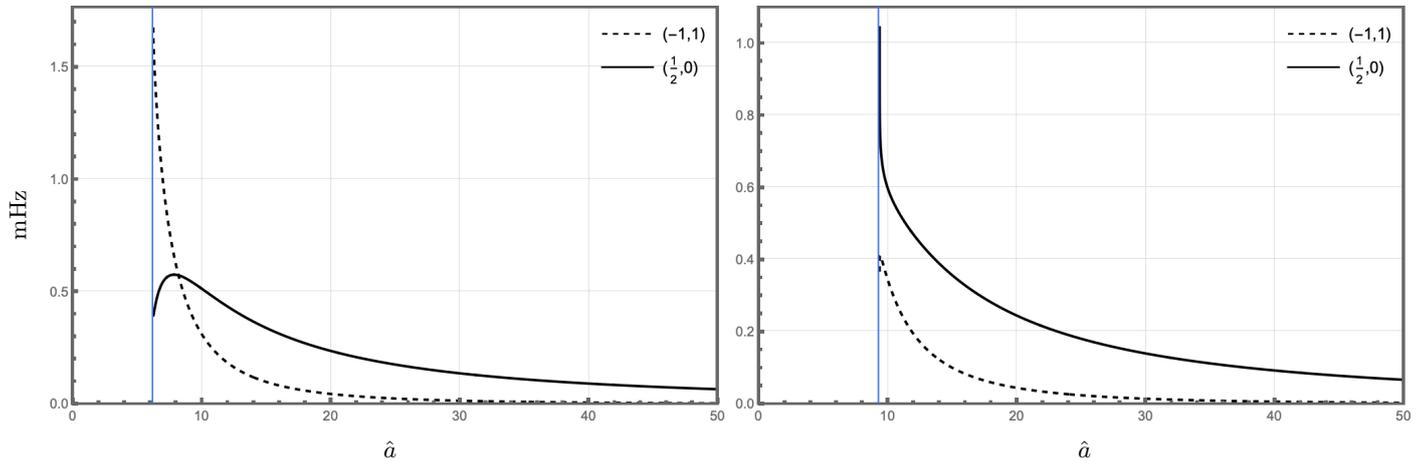

The distinction between whether the first resonance encountered by the inner binary arises from the electric or magnetic tidal moments originates from the nontrivial dependence of the outer orbit’s fundamental frequencies, $\Omega_{\hat{r}}$ and $\Omega_{\hat{\Psi}}$, on the semi-major axis $\hat{a}$ and eccentricity $\hat{e}$, as shown in the left panel of Fig.~\ref{fig_freq}.
The main consequence of this behavior is that in regions farther from the ISO of the SMBH, where the first precession resonance encountered by the inner binary is the magnetic $\left(-1,1\right)$ resonance, the system maintains a slightly higher eccentricity for a longer duration compared to regions closer to the ISO, where the first resonance is the electric $\left(\tfrac{1}{2},0\right)$.
This behavior is not restricted to the quasi-circular regime of the outer orbit (i.e., small values of $\hat{e}$), but also persists for finite outer eccentricities, as shown in Fig.~\ref{fig_freq}.

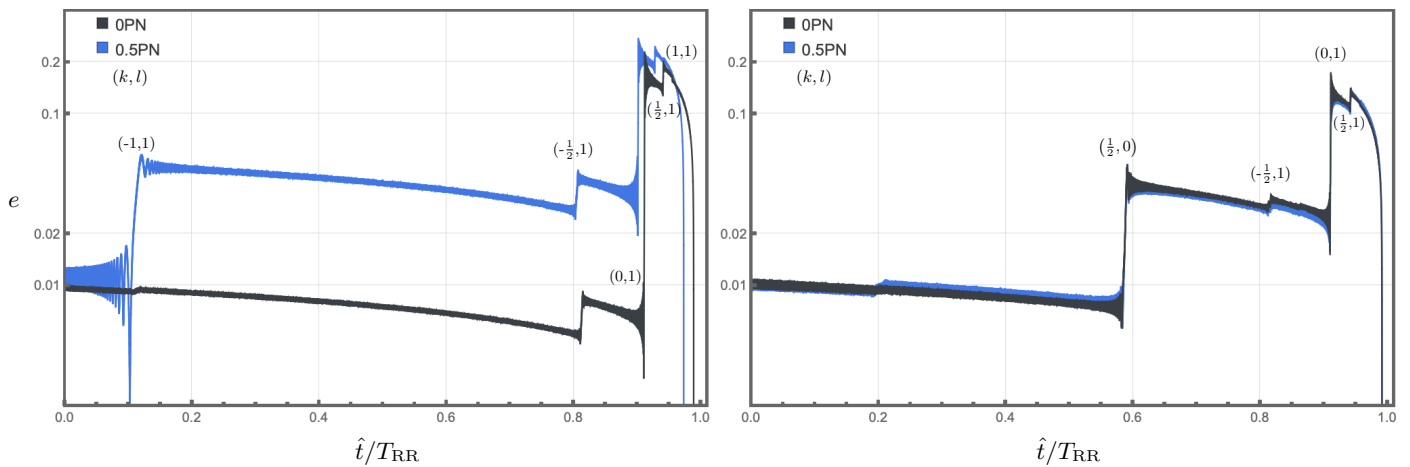
\begin{figure*}[ht]
    \centering
    \includegraphics[width=1\textwidth]{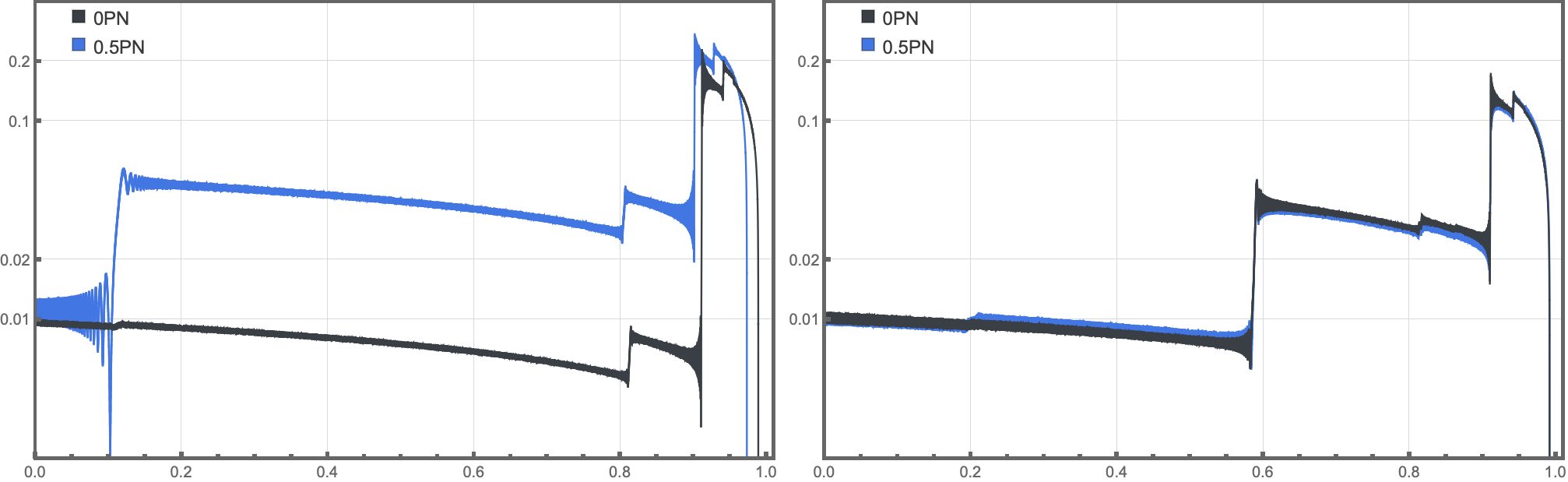}
    \begin{picture}(0,0)
        \put(-135,-2){\small $\hat t/T_{\rm in,RR}$  }
        \put(120,-2){\small $\hat t/T_{\rm in,RR}$  }
        \put(-265,93){$e$}
        \put(-226,140){\scalebox{0.7}{$(k,l)$}}
        \put(-224,115){\scalebox{0.65}{(-1,1)}}
        \put(-61,113){\scalebox{0.65}{(-$\frac{1}{2}$,1)}}
        \put(-40,65){\scalebox{0.65}{(0,1)}}
        \put(-26,128){\scalebox{0.65}{($\frac{1}{2}$,1)}}
        %\put(-120,105){\scalebox{0.65}{(-1,1)}} 
        \put(-19,150){\scalebox{0.65}{(1,1)}}

        \put(30,140){\scalebox{0.7}{$(k,l)$}}
        \put(143,114){\scalebox{0.65}{$\left(\frac{1}{2},0\right)$}}
        \put(200,104){\scalebox{0.65}{(-$\frac{1}{2}$,1)}}
        \put(224,149){\scalebox{0.65}{(0,1)}}
        \put(231,123){\scalebox{0.60}{($\frac{1}{2}$,1)}}
        %\put(38,114){\scalebox{0.65}{(-1,1)}} 
        %\put(237,150){\scalebox{0.65}{(1,1)}}
    \end{picture}
    \caption{Evolution of the eccentricity $e$ of the inner binary as a function of $\hat{t}/T_{\rm in,RR}$, including only the $0$PN tidal coupling (black curve) and both the $0$PN and $0.5$PN tidal couplings (blue curve), for a quasi-circular outer orbit.
    In the left panel, the initial mutual inclination between the inner and outer orbits is $I_0 \simeq 0^{\circ}$ (spin-aligned configuration), where precession resonances driven by the $0.5$PN magnetic coupling are enhanced.
    In contrast, in the right panel we set $I_0 = 60^{\circ}$, where all $0.5$PN magnetic contributions to precession resonances are suppressed.
    For both panels, the following parameters and initial conditions are used:
     $M_*=4\times 10^6\, \rm M_{\odot}$, $m_1=25\rm M_{\odot}$, $m_2=1.5\, \rm M_{\odot}$, $\hat e=0.05$, $\hat a=10 G M_*/c^2 \sim 0.4 \text{AU}$, $T_{\rm in,RR}\sim 6\,\text{yrs}$, GW peak frequency $f_{\mathrm{GW}}\sim 0.06\, \text{Hz}$, $a_0\sim 0.0003\, \text{AU}$ (close to first resonance), $e_0=0.01$,$\gamma_0=0^{\circ}$, $I_0=0.00001^{\circ}\sim0^{\circ}$ (left panel), $I_0=60^{\circ}$ (right panel), $\vartheta_0=0^{\circ}$, $\hat\Psi_0=0^{\circ}$.
}
\label{fig1_aligned}
\end{figure*}

The $0.5$PN resonant magnetic Hamiltonian, derived analytically in Eq.~\eqref{eq:H_B_Fourier_FirstOrder} at leading order in the outer-orbit eccentricity, depends sensitively on the mutual inclination $I$ between the inner and outer orbits.
To illustrate this dependence, Fig.~\ref{fig1_aligned} shows the evolution of the inner binary’s eccentricity for two different inclination angles in the quasi-circular outer-orbit regime.
In the left panel, we consider the spin-aligned configuration ($I \simeq 0^{\circ}$). In this case, the first precession resonance driven by the $0$PN electric tidal coupling, namely the $\left(\tfrac{1}{2},0\right)$ resonance listed in Table~\ref{tab:gamma_frequencies} is completely suppressed,\footnote{In agreement with the analytical model of Ref.~\cite{Cocco:2025adu}.} while the first precession resonance induced by the $0.5$PN magnetic coupling, the $\left(-1,1\right)$ resonance in Table~\ref{tab:gamma_frequencies}, plays a dominant role in shaping the eccentricity evolution.
In contrast, the right panel corresponds to $I = 60^{\circ}$, a configuration in which all $0.5$PN magnetic precession resonances (Table~\ref{tab:gamma_frequencies}) are suppressed for small values of the outer eccentricity $\hat{e}$. In this scenario, the inner binary experiences only the resonances associated with the $0$PN electric tidal coupling.
In both cases, the numerical results are in excellent agreement with the analytical predictions of Sec.~\ref{PerturbativeApproach}, which are valid to linear order in $\hat{e}$ (i.e., for nearly circular outer orbits).
Finally, we note that the dependence of the resonant Hamiltonian, both at $0$PN and $0.5$PN order, on the mutual inclination $I$ becomes increasingly intricate once higher-order terms in the outer eccentricity are included.
These higher-order effects can potentially trigger eccentricity jumps even at inclination angles for which precession resonances are suppressed in the quasi-circular limit.

\begin{figure*}[ht]
    \centering
    \includegraphics[width=1\textwidth]{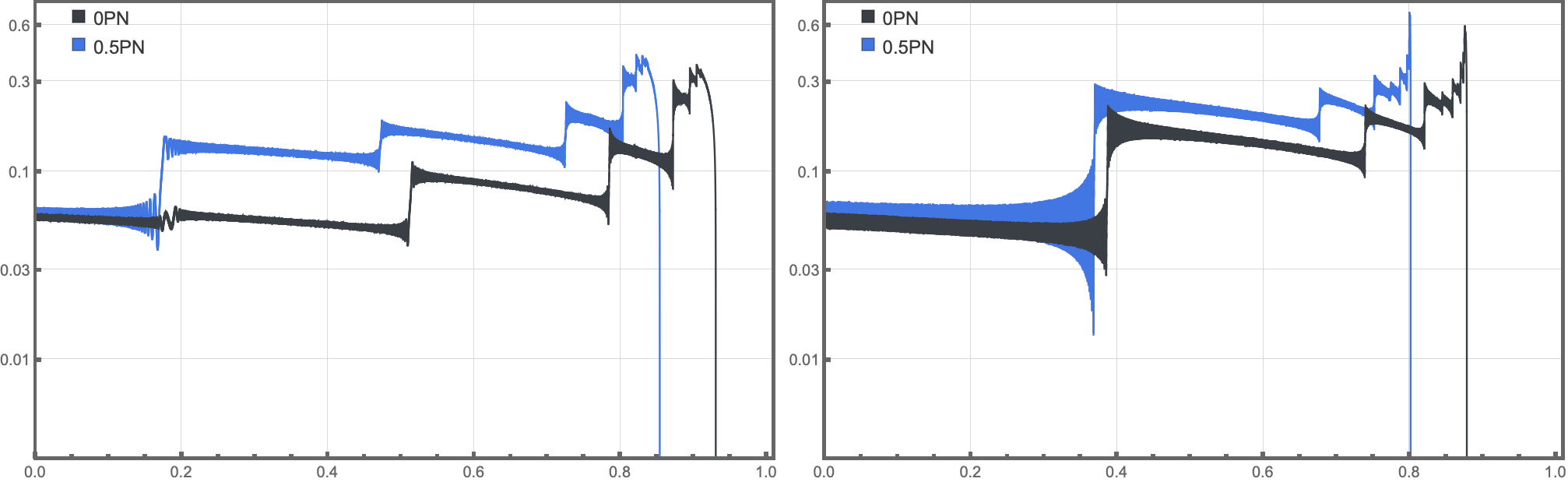}
    \begin{picture}(0,0)
        \put(-135,-2){\small $\hat t/T_{\rm in,RR}$  }
        \put(120,-2){\small $\hat t/T_{\rm in,RR}$  }
        \put(-265,93){$e$}
    \end{picture}
    \caption{Evolution of the eccentricity $e$ of the inner binary as a function of $\hat{t}/T_{\rm in,RR}$, including only the $0$PN tidal coupling (black curve) and both the $0$PN and $0.5$PN tidal couplings (blue curve). In the left panel, even for a moderate outer eccentricity of $\hat{e} = 0.2$, the $0$PN electric coupling in the Hamiltonian~\eqref{Hmarckaveinner_plusprecession} begins to contribute to precession resonances that, in the quasi-circular limit, arise solely from the $0.5$PN magnetic term. In the right panel, for a higher outer eccentricity ($\hat{e} = 0.5$), both the $0$PN electric and $0.5$PN magnetic couplings contribute to all precession resonances, making the two curves nearly indistinguishable. The only notable difference is that, with the $0.5$PN magnetic term included, the eccentricity jumps are slightly larger, leading to a faster inspiral of the inner binary.
    For both panels, the following parameters and initial conditions are used:
     $M_*=4\times 10^6\, \rm M_{\odot}$, $m_1=25\rm M_{\odot}$, $m_2=1.5\, \rm M_{\odot}$, $\hat e=0.2$ (left panel), $\hat e=0.5$ (right panel), $\hat a=10 G M_*/c^2 \sim 0.4 \text{AU}$ (left panel), $\hat a=15 G M_*/c^2 \sim 0.6 \text{AU}$ (right panel), $T_{\rm in,RR}\sim 6\,\text{yrs}$ (left panel), $T_{\rm in,RR}\sim 32\,\text{yrs}$ (right panel), GW peak frequency $f_{\mathrm{GW}}\sim 0.06\, \text{Hz}$ (left panel) and $f_{\mathrm{GW}}\sim 0.03\, \text{Hz}$ (right panel), $a_0\sim 0.0003\, \text{AU}$ (left panel) and $a_0\sim 0.0005\, \text{AU}$ (right panel), $e_0=0.06$, $\gamma_0=0^{\circ}$, $I_0=20^{\circ}$, $\vartheta=0^{\circ}$, $\hat\Psi=0^{\circ}$.
     The right panel corresponds to the most conservative configuration considered in this work, for which the separation between the inner orbital period and the tidal-variation timescale is reduced ($T_{\mathcal{E},{\rm per}}/T_{\rm in,orb},\, T_{\mathcal{B},{\rm per}}/T_{\rm in,orb} \sim 30$), but still exceeds our adopted threshold of one order of magnitude (see Sec.~\ref{subsec:validity_averaging}).
     }
\label{fig1_highhate}
\end{figure*}
Although our analytical model presented in Sec.~\ref{PerturbativeApproach} is valid only in the small-$\hat e$ regime, we can numerically integrate Eqs.~\eqref{LPEpeters} for finite values of the outer-orbit eccentricity.
This enables us to explore how precession resonances influence the evolution of the inner binary even for large $\hat e$.
As shown in Fig.~\ref{fig1_highhate}, increasing the outer eccentricity enhances higher-order corrections in $\hat e$ that contribute to each $\left(k,l\right)$ precession resonance through both the $0$PN electric and the $0.5$PN magnetic couplings.
In the left panel, we find that already at $\hat e = 0.2$, the $0$PN electric tidal coupling begins to contribute to the $\left(-1,1\right)$ resonance, which in the quasi-circular limit arises solely from the $0.5$PN magnetic coupling.
This trend becomes even more pronounced in the right panel, where for $\hat e = 0.5$ both tidal couplings contribute to all $\left(k,l\right)$ precession resonances.
Note that in this right panel, the analytic resonance dictionary derived in Sec.~\ref{sec:analytic} for $\hat e \ll 1$ is being extrapolated beyond its strict regime of validity.
The primary distinction between the two cases is that the $0.5$PN magnetic coupling amplifies the eccentricity jumps of the inner binary, leading to higher peak eccentricities and a slightly faster inspiral toward merger.

Finally, each eccentricity jump induced by a precession resonance during the evolution of the inner binary produces a corresponding shift in the phase of its gravitational waveform~\cite{Samsing:2024syt,Hendriks:2024gpp,Bonga_et_al_2019,Gupta_et_al_2021,Flanagan_Hughes_Ruangsri,Chandramouli_Yunes,Flanagan_Hinderer_2012,Zwick:2025ine}.
This effect is particularly relevant for our analysis, since in all the scenarios presented in this section the peak frequency of the GWs emitted by the inner binary, given by~\cite{Wen_2003}
\begin{equation} \label{eq:f_Wen}
    f_{\mathrm{GW}} \approx \frac{\sqrt{G M}}{\pi\left[a\left(1-e^2\right)\right]^{3 / 2}}(1+e)^{1.1954} \,,
\end{equation}
remains entirely within the LISA sensitivity band ($0.001~\mathrm{Hz}$ -- $0.1~\mathrm{Hz}$) throughout the binary’s evolution.

To obtain a qualitative estimate of the dephasing induced by the magnetic resonances, we compute the accumulated number of GW cycles \cite{Kavanagh:2020cfn, Li:2026uva,Maselli:2020zgv} 
\begin{equation} \label{eq:N_GW}
    N_{\text {cycles }} \equiv \int_{\hat{t}_{\rm i}}^{ \hat{t}_{\rm f}} f_{\mathrm{GW}}(t) ~ d  t~,
\end{equation} 
where we use $f_{\rm GW}$ as defined in Eq.~\eqref{eq:f_Wen}.
Having this in hand, we look at the difference between the evolution including the $0.5$PN coupling and the one including only the $0$PN coupling, namely $\Delta N \equiv N_{0.5\rm PN}-N_{0\rm PN}$. 
The integration is performed from $\hat{t}_i=0$ to $ \hat{t}_f=0.95~T_{0.5\rm PN}$, where $T_{0.5\rm PN}$ is the merger time corresponding to the magnetic case, which is the shorter one. This conservative cutoff enables us to compare the accumulated cycles over the same time interval while excluding the late inspiral, for which our model becomes less reliable. 
Note that this integrated estimate mixes the localized dephasing produced at resonance crossings with the cumulative difference in the inspiral rate, since the evolution including the magnetic coupling reaches merger earlier.

\begin{table}[ht]
\centering
\caption{Difference in the accumulated number of GW cycles between the $0.5$PN and $0$PN evolutions.}
\begin{tabular*}{\linewidth}{
  @{\extracolsep{\fill}}
  c
  c
  c
  @{}
}
\toprule
Figure & Panel & $\Delta N$ \\
\midrule
Fig.~$1$ & Left  & $\approx 3 \times 10^5$ \\
       & Right & $\approx 5 \times 10^5$ \\[1ex]
Fig.~$3$ & Left  & $\approx 7\times 10^5$ \\
       & Right & $ \approx 5 \times 10^4$ \\[1ex]
Fig.~$4$ & Left  & $ \approx 2 \times 10^6$ \\
       & Right & $ \approx 4\times 10^6$ \\
\bottomrule
\end{tabular*}
\label{tab:N_cycles_ratios}
\end{table}
The results summarized in Tab.~\ref{tab:N_cycles_ratios} show that magnetic tidal coupling leads to an increase in the accumulated number of GW cycles in every configuration we considered.
The difference becomes more pronounced as the outer eccentricity grows, consistently with the stronger modulation of the orbital evolution.
Adopting the conservative detectability criterion that a phase difference of order one radian is measurable \cite{Li:2026uva,Maselli:2020zgv}, our qualitative estimate suggests that the magnetic tidal coupling may produce a potentially observable dephasing.
This back-of-the-envelope analysis motivates the development of dedicated waveform models to properly quantify the detectability of relativistic effects arising from the $0$PN electric and $0.5$PN magnetic tidal couplings.
In particular, the magnetic coupling, having no Newtonian ($0$PN) counterpart, could provide a distinctive relativistic signature of the SMBH’s tidal field in the strong-gravity regime.

\section{Conclusions} \label{sec:CONCLUSIONS}

In this paper, we have extended previous analyses of precession resonances in hierarchical three-body systems~\cite{Kuntz:2021hhm,Cocco:2025adu}
to include a quadrupole magnetic tidal coupling to the inner binary. 
This coupling is crucial for accurately describing strong-gravity interactions between a compact-object binary and a SMBH.

We have shown that precession resonances arise from a $0.5$PN coupling to the magnetic tidal field, and that they satisfy the resonance condition~\eqref{eq:intro} with $q=1$, which
%\sout{, in the small outer eccentricity limit,} 
leads to additional jumps in the eccentricity of the inner binary compared to what was found in Ref.~\cite{Cocco:2025adu}.
Furthermore, our numerical analysis suggests that the inclusion of magnetic tidal effects enhances the magnitude of these eccentricity jumps, potentially leading to an accelerated merger of the binary components.

An important extension of this work would be to construct gravitational waveforms for the inner binary that explicitly include the effects of precession resonances. This could be achieved by following the formalism of Ref.~\cite{Santos:2025ass}, in which the inner binary of a b-EMRI system is modeled as a point particle endowed with a multipolar structure. To incorporate precession resonances within this framework, one needs to account for the tidal interaction between the two companions in the b-EMRI in the evolution of the multipoles characterizing the point particle.

Moreover, this analysis could be extended by modeling the SMBH with the Kerr metric, thereby accounting for its spin.
Such an extension would likely yield a more general resonance condition that includes the additional fundamental frequency associated with the polar angle $\hat{\theta}$ of the outer orbit.
This refinement would be essential for investigating how the SMBH’s spin influences the evolution of the inner binary and the resulting gravitational-wave signatures.

Finally, it would be interesting to explore whether the inclusion of the $0.5$PN magnetic tidal coupling influences secular phenomena such as the ZLK mechanism. 
In particular, for an inner binary subject to a strong-gravity tidal field, the magnetic tidal moments could provide a measurable contribution to the secular evolution if the binary’s dynamics is sufficiently relativistic.

\section*{Acknowledgments}
We thank the anonymous referee for carefully reading the manuscript and for the constructive comments and suggestions provided.
We thank L. Rezzolla for useful feedback and stimulating discussions which led to Appendix D. We thank F. Camilloni, A. Kuntz, João S. Santos and F. Marzocco for insightful feedback.
M. Cocco, G. Grignani, M. Orselli, D. Panella and D. Pica acknowledge financial support from the Italian Ministry of University and Research (MUR) through the program ``Dipartimenti di Eccellenza 2023-2027" (Grant SUPER-C). G. Grignani and M. Orselli also acknowledge support from ``Fondo di Ricerca d'Ateneo"  2023 (GraMB) of the University of Perugia. G. Grignani, M. Orselli and D. Pica acknowledge support from the Italian Ministry of University and Research (MUR) via the PRIN 2022ZHYFA2, GRavitational wavEform models for coalescing compAct binaries with eccenTricity (GREAT). M. Cocco, T. Harmark, M. Orselli and D. Panella acknowledge support by the “Center of Gravity”, which is a Center of Excellence funded by the Danish National Research Foundation under grant no. DNRF184. D. Pica thanks Goethe University (Frankfurt) for hospitality.

\appendix

\section{Marck's tetrad and accelerated tetrad} \label{app.tetrads}
\renewcommand{\theequation}{\thesection.\arabic{equation}}

The local inertial frame used in our analysis is defined by Marck’s tetrad $\bar{\lambda}_{A}^{\mu}$, first derived in Ref.~\cite{Marck:1983} for the Kerr spacetime. It provides an orthonormal frame that is parallel transported along a geodesic. The tetrad is characterized by vanishing four-acceleration and angular velocity, namely
\begin{equation}
    a^i \equiv \bar{\lambda}^i_\mu\frac{D\bar{\lambda}^\mu_{0}}{d\hat{\tau}}=0 \,,
    ~~~
    {\omega}^i\equiv-\frac{1}{2}{\epsilon^{ij}}_{k}{\bar{\lambda}}^\mu_j\frac{D{\bar{\lambda}}_\mu^k}{d\hat{\tau}}=0 \,,
\end{equation}
and can be schematically written as 
\begin{equation}
\begin{aligned} 
    \label{Marck's_tetrad}
    \bar{\lambda}_{0}^\mu & \equiv \frac{u^\mu}{c}= e_{0}^\mu \,, \\
    \bar{\lambda}_{1}^\mu & = \cos{\hat{\Psi}} e_{1}^\mu- \sin{\hat{\Psi}} e_{2}^\mu \,, \\
    \bar{\lambda}_{2}^\mu & =\sin{\hat{\Psi}} e_{1}^\mu+ \cos{\hat{\Psi}} e_{2}^\mu \,, \\
    \bar{\lambda}_{3}^\mu & =e_{3}^\mu \,.
\end{aligned}
\end{equation}
Here, $u^{\mu}$ denotes the four-velocity along the geodesic around which the local inertial frame is built, $e_A^{\mu}$ represents the rotating frame with a non-zero angular velocity, built by Marck using the Killing-Yano tensor of the Kerr spacetime \cite{Marck:1983}. The quantity $\hat{\Psi}$ is Marck's angle \cite{Marck:1983}, which ensures that Marck's tetrad is parallel transported along the geodesic. The explicit components of this tetrad can be found in Refs.~\cite{Marck:1983, Camilloni_2023}, to which we refer the reader.\footnote{The Schwarzschild limit is recovered by setting the black hole spin to zero.}

When the inner binary is modeled at the $0$PN order, the dynamics of its CoM and its relative motion are independent, allowing us to assume that the CoM follows a geodesic around the SMBH~\cite{Camilloni_2024,maeda2023chaoticvonzeipellidovkozaioscillations,Maeda:2025row}. 
However, once $0.5$PN corrections are included, the CoM and the relative motions of the inner binary become coupled, as can be seen explicitly from Eq.~\eqref{L1/2interactionterm}.
Nevertheless, as pointed out in Ref.~\cite{Maeda_2023}, it remains possible to decouple the two by introducing an acceleration $\boldsymbol{a}$ that cancels the $0.5$PN interaction term~\eqref{L1/2interactionterm} between the CoM and the relative motion.
To this end, we recall that the equations of motion of a system are unchanged by the addition of a total time derivative to the Lagrangian. Exploiting this freedom, and following Ref.~\cite{Maeda_2023}, we can rewrite Eq.~\eqref{L1/2interactionterm} as
\begin{equation}
    \label{newL05int}
    L_{\rm 0.5 \text{-int}}= 2 \mu c^2 \left[\frac{1}{3 c} \frac{d \left(\epsilon_{ijk} \mathcal{B}^{j}{}_{\ell}\right)}{d \hat{\tau}} x^{k} x^{\ell} + \epsilon_{ijk} \mathcal{B}^{j}{}_{\ell} x^{k} \frac{v^{\ell}}{c}\right] X^{i} \,,
\end{equation}
from which we can define an acceleration given by 
\begin{equation}
    \label{a_i}
    a_{i} = \frac{2 \mu c^2}{m_1 + m_2} \left[\frac{1}{3 c} \frac{d \left(\epsilon_{ijk} \mathcal{B}^{j}{}_{\ell}\right)}{d \hat{\tau}} x^{k} x^{\ell} + \epsilon_{ijk} \mathcal{B}^{j}{}_{\ell} x^{k} \frac{v^{\ell}}{c}\right] \,.
\end{equation}
In an accelerated reference frame characterized by the acceleration~\eqref{a_i}, the CoM contribution to the Lagrangian, $L_{a\text{-CoM}}$ in Eq.~\eqref{eq:L_CoM_a}, which accounts for fictitious forces, exactly cancels the $0.5$PN interaction between the CoM dynamics and the relative motion of the inner binary. This allows the two motions to be treated as fully decoupled.

Note that the first term in the acceleration~\eqref{a_i} represents the time derivative of a quadrupole tidal moment and therefore scales as an octupolar contribution~\cite{PoissonVlasov}, which we neglect in our analysis.
The second term corresponds to the gravitoelectromagnetic force, as discussed in Ref.~\cite{Chen_2022}.
Although working in an accelerated reference frame with acceleration given by Eq.~\eqref{a_i} allows us to decouple the CoM motion of the inner binary from its relative motion, the CoM, strictly speaking, no longer follows a geodesic around the SMBH.
Consequently, the reference frame in which we describe the evolution of the inner binary is not strictly inertial, and, in principle, Marck’s tetrad cannot be directly used to analyze its dynamics.
Nevertheless, as we show below, within the quadrupolar order of the small-tide approximation, the system can still be effectively described by assuming geodesic motion for the CoM up to $0.5$PN order.

In detail, the geodesic equations in the accelerated reference frame can be written as
\begin{equation} 
    \label{duCoM}
    \frac{D u_{\mathrm{CoM}}^\mu}{d \hat{\tau}}=a^\mu=e^{\mu i} a_{i} \,,
\end{equation}
where $a_i$ is defined in Eq.~\eqref{a_i}, and $u^{\mu}_{\rm CoM}$ denotes the four-velocity of the inner binary’s CoM, which can be written as
\begin{equation}
\label{eq:fourv}
u^\mu_{\rm CoM}=u_{(0)}^\mu+u_{(1)}^\mu \,.
\end{equation}
Here, $u_{(0)}^{\mu} \equiv c\, e_0^{\mu}$ denotes the four-velocity in the rotating frame, i.e., that associated with the geodesic motion of the CoM, while $u_{(1)}^{\mu}$ represents the correction to the four-velocity arising from the acceleration~\eqref{a_i}.
The latter can be expressed in terms of the deviation $X^{\mu}_{(1)}$ from the geodesic trajectory as
\begin{equation}
u_{(1)}^\mu=\frac{d {X}_{(1)}^\mu}{d \hat{\tau}} \,,
\end{equation}
where 
\begin{equation}
\label{eq:coordinatesdev}
{X}_{(1)}^\mu=\left(c \hat{t}_{(1)}, \hat{r}_{(1)}, \hat{\theta}_{(1)}, \hat{\phi}_{(1)}\right) \,.
\end{equation}

Note that we assume $X_{(1)}^{\mu}$ and $u_{(1)}^{\mu}$ to be small perturbations. This is consistent with the small-tide approximation, which allows the interaction between the inner binary and the SMBH to be treated as weak.
Since the acceleration~\eqref{a_i}, which gives rise to the deviations $X_{(1)}^{\mu}$ and $u_{(1)}^{\mu}$ from geodesic motion, is defined in terms of the quadrupolar magnetic tidal moments $\mathcal{B}_{ij}$, nonlinear terms in Eq.~\eqref{duCoM} would correspond to corrections beyond the quadrupole order.

Retaining only linear terms in the deviation from geodesic motion, Eq.~\eqref{duCoM} can be written as
\begin{equation}\label{geodev}
\frac{d u_{(1)}^\mu}{d \hat{\tau}}+2 \Gamma_{\rho \sigma}^\mu u_{(0)}^\rho u_{(1)}^\sigma+\frac{\partial \Gamma_{\rho \sigma}^\mu}{\partial X^{\alpha}} X_{(1)}^{\alpha} u_{(0)}^\rho u_{(0)}^\sigma=a^\mu \,.
\end{equation}
This represents a set of coupled, second-order differential equations for the coordinates defined in Eq.~\eqref{eq:coordinatesdev}, whose explicit form is given in Ref.~\cite{Maeda_2023}.
Without further assumptions, this system can only be solved numerically, as demonstrated for circular orbits around a Schwarzschild black hole in Ref.~\cite{Maeda_2023}, for Kerr black holes in Refs.~\cite{maeda2023chaoticvonzeipellidovkozaioscillations,Maeda:2025row}, and within the gravitoelectromagnetic framework for generic eccentric orbits in Ref.~\cite{Chen_2022}. 
In all the works mentioned above, it has been shown that these deviations remain small compared to the $0$PN geodesic motion.
However, it is important to assess whether they affect the tetrad~\eqref{Marck's_tetrad}, which is used to construct the quadrupole tidal moments~\eqref{Weylcomp}.
To this end, we note that the acceleration~\eqref{a_i} is already of quadrupolar order in the tidal multipole expansion, and therefore induces quadrupolar corrections to the four-velocity~\eqref{eq:fourv} and, consequently, to the tetrad~\eqref{Marck's_tetrad}.
When these corrections are contracted with the Weyl tensor in the computation of the tidal moments~\eqref{Weylcomp}, they yield contributions beyond the quadrupole order, which can be safely neglected in our analysis.
Without explicitly solving the system~\eqref{geodev}, and assuming $\hat{r}$ constant,\footnote{This is a good approximation in our regime, since for small outer eccentricities $\hat e \ll 1$ the radial coordinate $\hat r$ varies negligibly during the evolution of the inner binary.} we verified that the corrections to the tetrad~\eqref{Marck's_tetrad} arising from the acceleration~\eqref{a_i} are indeed of quadrupolar order and therefore negligible.
Ref.~\cite{Maeda_2023} reached the same conclusion by considering a circular outer orbit and small deviations from purely radial motion.
Guided by this reasoning, we describe the evolution of the inner binary using the local inertial frame defined by Marck’s tetrad, as if its CoM continues to move on a geodesic around the Schwarzschild SMBH.

Deviations from geodesic motion would modify the conserved quantities of the outer orbit, such as the energy $\hat{E}$ and angular momentum $\hat{L}$, by introducing corrections of quadrupolar order.
These corrections can safely be neglected, since the constants of the outer motion influence the inner relative motion only through tidal couplings, which are themselves already of quadrupolar order.

\section{Tidal moments for a Schwarzschild geodesic} \label{Tidal_Moments_Schwarzschild}

We consider a SMBH described by the Schwarzschild metric, given in Eq.~\eqref{sch_metric}.
In the local inertial frame defined by Marck’s tetrad, the tidal moments can be computed directly from Eqs.~\eqref{Weylcomp} and~\eqref{Tidaldef}.
The electric tidal tensor $\mathcal{E}_{ij}$ then takes the explicit form
\begin{equation} 
    \label{EijMarck}
   \small \mathcal{E}_{i j}=\frac{\hat{r}_s}{\hat{r}^5}\left(\begin{array}{ccc}
\frac{\hat{r}^2-3\left(\hat{L}^2+\hat{r}^2\right) \cos ^2\hat{\Psi}}{2} & -\frac{3 \left(\hat{L}^2+\hat{r}^2\right) \sin 2 \hat{\Psi}}{4} & 0 \\
-\frac{3\left(\hat{L}^2+\hat{r}^2\right) \sin 2 \hat{\Psi}}{4} & \frac{\hat{r}^2-3\left(\hat{L}^2+\hat{r}^2\right) \sin ^2\hat{\Psi}}{2} & 0 \\
0 & 0 & \frac{3 \hat{L}^2+\hat{r}^2}{2 }
\end{array}\right) \,,
\end{equation}
while the magnetic tidal tensor $\mathcal{B}_{i j}$ is
\begin{equation} 
    \label{BijMarck} 
     \mathcal{B}_{i j}=\frac{3 \hat{r}_s \hat{L} \sqrt{\hat{L}^2+\hat{r}^2}}{2 \hat{r}^5} \left(\begin{array}{ccc}
        0 & 0 & -\cos \hat{\Psi} \\
        0 & 0 &  -\sin \hat{\Psi} \\
         -\cos \hat{\Psi} & -\sin \hat{\Psi} & 0
\end{array}\right) \,.
\end{equation}

\section{Distant-star frame} \label{App:DistantStarFrame}

The Hamiltonian in Eq.~\eqref{HbinaryEqBq} provides a local description of the inner binary dynamics.
In this Appendix, we reformulate the problem from the perspective of an asymptotic observer by introducing a non-inertial reference frame with zero acceleration but nonzero angular velocity $\Omega_{\rm g}$, known as the \textit{distant-star frame}~\cite{Bini_2016}.
This frame allows us to describe the evolution of the inner binary as seen by an observer at spatial infinity.

Starting from Marck’s tetrad, the distant-star frame can be obtained through a spatial rotation of the form $e_{\mu}^i = R_{j}^{\ i} \bar{\lambda}_{\mu}^{j}$, where $R^{\ i}_{j}$ is a time-dependent rotation matrix that relates the locally inertial (Marck's) frame $\bar{\lambda}_{\mu}^{j}$ to the rotating distant-star frame $e_{\mu}^i$. The rotation matrix is given by
\begin{equation} \label{rotationmatrix}
R_j^i=\left(\begin{array}{ccc}
\cos \Delta(\hat{\tau}) & -\sin \Delta(\hat{\tau}) & 0 \\
\sin\Delta(\hat{\tau}) & \cos \Delta(\hat{\tau}) & 0 \\
0 & 0 & 1
\end{array}\right),
\end{equation}
where the accumulated rotation angle is
\begin{equation} 
    \label{phasedelta}
    \Delta(\hat{\tau})=\int d\hat{\tau} ~\Omega_{\rm g} \,. 
\end{equation}
Here, the quantity
\begin{equation}
\Omega_{\rm g}=\frac{d \hat{\phi}}{d\hat{\tau}}-\frac{d \hat{\Psi}}{d\hat{\tau}}
\end{equation}
represents the angular velocity of the rotating tetrad $e_{\mu}^i$ with respect to the locally inertial (Marck's) frame. It is expressed in terms of the azimuthal coordinate $\hat{\phi}$ of the external Schwarzschild spacetime and Marck’s angle $\hat{\Psi}$, which encodes the parallel transport of the tetrad along the geodesic.  

The coordinates associated with the distant-star frame $\boldsymbol{r}$ are given by
\begin{equation}
    r^i=R^{i}_{j}x^j \,,
\end{equation}
and consequently, the velocities $\boldsymbol{\nu}$ read
\begin{equation}
        \nu^i =R_j^i \dot{x}^j+\dot{R}_j^i x^j=R_j^i v^j+\left(\boldsymbol{\Omega_{\rm g}} \times \boldsymbol{r}\right)^i \,.
\end{equation}
This relation can be inverted to express the velocities $\boldsymbol{v}$ measured in the local inertial (Marck's) frame in terms of the velocities $\boldsymbol{\nu}$ defined in the distant-star frame, namely
\begin{equation}
v^i=\left(R^{-1}\right)_j^i\left(\boldsymbol{\nu}-\boldsymbol{\Omega_{\rm g}}\times \boldsymbol{r}\right)^j \,,
\end{equation}
where $R^{-1}$ is the inverse of the rotation matrix defined in Eq.~\eqref{rotationmatrix}.

By introducing the rotated magnetic tidal components $\beta^{\rm q}_i=R^j_i\mathcal{B}^{\rm q}_j$, the Lagrangian in the distant-star frame can be written as
\begin{equation}
\label{eq:lag_dist}
\begin{aligned}
        L_{\text{inner}}^* =&\frac{1}{2} \mu\left(\boldsymbol{\nu}-\boldsymbol{\Omega_{\rm g}} \times \boldsymbol{r}\right)^2+\frac{G M \mu}{r}-\frac{\mu c^2 r^2}{2}\mathcal{E}_{\rm q}\\
        &-\left(R^{-1}\right)^i _j\left(\boldsymbol{\nu}-\boldsymbol{\Omega_{\rm g}} \times \boldsymbol{r}\right)^j\left(R^{-1}\right)_i^m \beta^{\rm q}_m
\end{aligned}
\end{equation}
From the Lagrangian~\eqref{eq:lag_dist}, the canonical momentum is obtained as
\begin{equation} 
    \label{canonicalmomentumdistant}
    \boldsymbol{\pi}=\frac{\partial L_{\text{inner}}^*}{\partial \boldsymbol{\nu}}=\mu(\boldsymbol{\nu}-\boldsymbol{\Omega_{\rm g}} \times \boldsymbol{r})-\boldsymbol{\beta^{\rm q}} \,,
\end{equation}
from which the corresponding Hamiltonian follows by performing a Legendre transformation of $L_{\text{inner}}^*$
\begin{equation}
H_{\text{inner}}^*=\boldsymbol{\pi}\cdot\boldsymbol{\nu}-L_{\text{inner}}^* \,.
\end{equation}
Using 
\begin{equation}
\boldsymbol{\nu}=\frac{\boldsymbol{\pi}+\boldsymbol{\beta^{\rm q}}}{\mu}+\boldsymbol{\Omega_{\rm g}} \times\boldsymbol{r} \,,
\end{equation}
we get
\begin{equation}
    H_{\text{inner}}^*= \frac{ \boldsymbol{\pi}^2}{2 \mu}+\frac{\boldsymbol{\pi} \cdot \boldsymbol{\beta^{\rm q}}}{\mu}+\frac{{\boldsymbol{\beta^{{\rm q}}}}^2}{2 \mu}-\frac{G M \mu}{r}+\frac{\mu c^2 r^2}{2}  \mathcal{E}_{\rm q}  +\boldsymbol{\pi} \cdot\left(\boldsymbol{\Omega_{\rm g}} \times \boldsymbol{r}\right)~.
\end{equation}
Retaining terms only up to the quadrupolar order, this expression can be recast as
\begin{equation} 
    \label{Hdistant}
    H_{\text{inner}}^*= H_{\text{inner}}+\boldsymbol{\Omega_{\rm g}}\cdot\boldsymbol{L_{\text{in}}} \,.
\end{equation}
Here, $H_{\rm inner}$ denotes the Hamiltonian obtained in Marck’s (i.e., locally inertial) frame, while the last term in Eq.~\eqref{Hdistant} represents the {\it gyroscope precession}~\cite{Bini_2016,Camilloni_2024},
This term accounts for the precession of the inner binary’s orbital angular momentum, $\boldsymbol{L_{\text{in}}} = \boldsymbol{r} \times \boldsymbol{\pi}$, induced by the curvature of the external spacetime in which the system evolves.
 It is worth emphasizing that gyroscope precession is a global effect, and therefore cannot be analyzed directly within a purely local inertial frame~\cite{Camilloni_2024}.

We note that the two Hamiltonians defined in Eqs.~\eqref{HbinaryEqBq} and~\eqref{Hdistant} are related through a canonical transformation of the second kind~\cite{goldstein:mechanics}, characterized by the generating function
\begin{equation} 
    \label{GeneratingFunction}
    F_2(q, P, \hat{\tau})=R_j^i x^j \pi_i \,,
\end{equation}
where
\begin{equation}
    \begin{array}{ll}
        q_i=x_i & Q_i=r_i \,, \\
        p_i=p_i & P_i=\pi_i \,.
    \end{array}
\end{equation}
This consistently gives
\begin{equation}
    p_i =\frac{\partial F_2}{\partial x_i}\,, \qquad r_i  =\frac{\partial F_2}{\partial \pi_i}\,,
\end{equation}
and
\begin{equation}
\begin{aligned}
H_{\text{inner}}^* &= H_{\text{inner}} + \partial_{\hat{\tau}} F_2 \\
&= H_{\text{inner}} + \dot{R}^{\,i}{}_j\, x^j \pi_i \\
&= H_{\text{inner}} + \bm{\Omega}_{\rm g}\cdot \bm{L}_{\text{in}} \, .
\end{aligned}
\end{equation}

\section{Environmental effects} \label{App:Environmental effects}
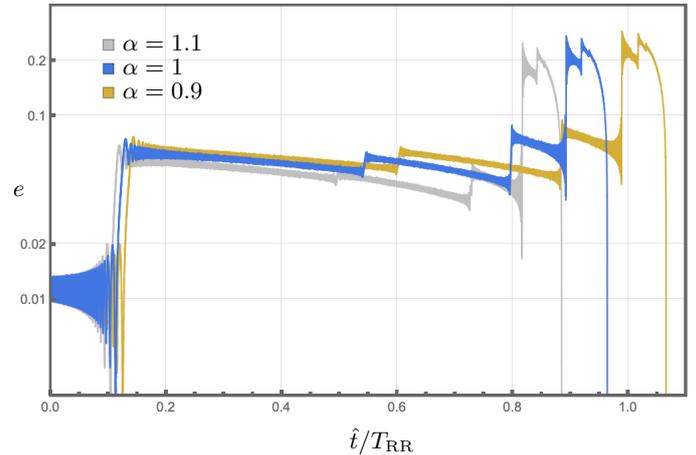
\begin{figure}[ht]
    \centering
    \includegraphics[width=0.49\textwidth]{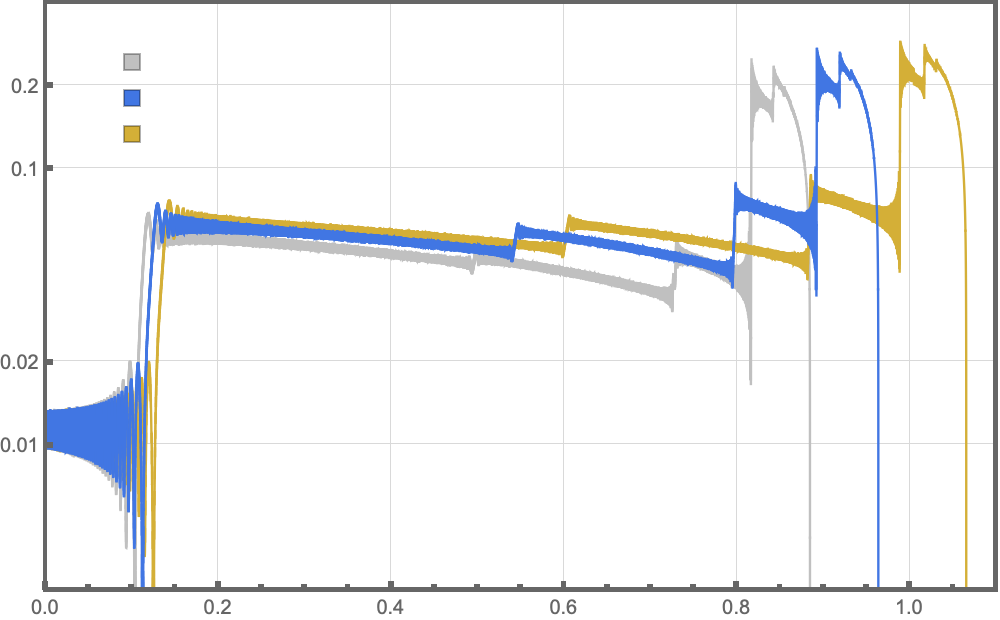}
    \begin{picture}(0,0)
        \put(0,-2){\small $\hat t/T_{\rm in,RR}$  }
        \put(-85,139){\small $\alpha =1$ }
         \put(-85,148){\small $\alpha =1.1$ }
         \put(-85,130){\small $\alpha =0.9$ }
        \put(-126,93){$e$}
    \end{picture}
    \caption{Evolution of the eccentricity $e$ of the inner binary for different values of the parameter $\alpha$. The following parameters and initial conditions are used:
     $M_*=4\times 10^6\, \rm M_{\odot}$, $m_1=25\rm M_{\odot}$, $m_2=1.5\, \rm M_{\odot}$, $\hat e=0.08$, $\hat a=10 G M_*/c^2 \sim 0.4 \text{AU}$, RR timescale as defined in Eq.~\eqref{T_RR_approx} $T_{\rm in,RR}\sim 6\,\text{yrs}$, GW peak frequency $f_{\mathrm{GW}}\sim 0.06\, \text{Hz}$, $a_0\sim 0.0003\, \text{AU}$ (close to first resonance), $e_0=0.01$, $\gamma_0=0^{\circ}$, $I_0=15^{\circ}$, $\vartheta_0=0^{\circ}$, $\hat\Psi_0=0^{\circ}$.}
\label{fig:dissipative}
\end{figure}

In our analysis, the inner binary is evolved through multiple precession resonances by including only the dissipative effects associated with gravitational-wave RR, as described by Eqs.~\eqref{PetersEQ}. However, additional environmental effects, such as the presence of gas, stars, or a dark matter halo, may also introduce dissipative contributions to the binary’s evolution~\cite{Cardoso:2020iji, Zwick:2021ayf, CanevaSantoro:2023aol, Ishibashi:2024wwu, Barausse:2007dy, Barausse:2014tra, Cardoso:2020lxx, Cole:2022yzw, Chen:2025qyj}. To obtain a qualitative estimate of how such effects could modify the evolution and resonant dynamics of the inner binary, we introduce a dimensionless multiplicative parameter $\alpha$ in Peters’ equations, namely
\begin{equation}
    \begin{split} 
    \label{PetersEQ_alpha}
    \frac{d a}{d t}\Bigg|_{\mathrm{GW}}&=-\alpha\frac{64}{5} \frac{G^3 \mu~ M^2}{c^5 a^3} \frac{1}{\left(1-e^2\right)^{7 / 2}} \left(1+\frac{73}{24} e^2+\frac{37}{96} e^4\right), \\
    \frac{d e}{d t}\Bigg|_{\mathrm{GW}}&=-\alpha\frac{304}{15} \frac{G^3 \mu ~ M^2}{c^5 a^4} \frac{e}{\left(1-e^2\right)^{5 / 2}}\left(1+\frac{121}{304} e^2\right) .
\end{split}
\end{equation}
By setting $\alpha=1$, we recover the analysis presented in the main text, where the only dissipative effects arise from RR (blue curve in Fig.~\ref{fig:dissipative}). When $\alpha>1$, we assume that environmental effects enhance the dissipative contributions to the inner binary's dynamics (silver curve in Fig.~\ref{fig:dissipative}). For completeness, we also consider the case $\alpha<1$, representing a scenario in which the efficiency of dissipative effects is suppressed (gold curve in Fig.~\ref{fig:dissipative}). 
In Fig.~\ref{fig:dissipative}, for the silver and gold curves, we consider a deviation of $10\%$ from the value $\alpha=1$ because dissipative contributions from environmental effects are typically smaller than the RR contributions described by Peters' equations \cite{Barausse:2007dy,Barausse:2014tra}. Furthermore, Fig.~\ref{fig:dissipative} shows that including environmental effects in the dissipative terms of our model affects only the timing at which the resonant conditions are met. However, the number of precession resonances encountered by the inner binary system during its evolution and their amplitude remain unchanged. This is consistent with the fact that, in our model, dissipative effects do not enter the resonant Hamiltonian but serve only as drivers for the evolution of the inner binary system.
This environmental model, however, does not include effects that would have an impact on the CoM motion, as we also neglect other dissipative effects of the outer motion, such as the outer RR.

%%%%%%%%%%%%%%%%%%%%%%%%%%
%%%%%%%%%%%%%%%%%%%%%%%%

\appendix

%%%%%%%%%%%%%%%%%%%%%%%%%%%%%%%%%%%%%%%%%%%

\newpage
\bibliographystyle{apsrev4-1}
\bibliography{sample}

\end{document}